\documentclass[%
 reprint,
 amsmath,amssymb,
 aps,
]{revtex4-2}

\usepackage{graphicx}
\usepackage{amsmath}
\usepackage{amsfonts}
\usepackage{algorithmic}
\usepackage{amssymb}
\usepackage{textcomp}
\usepackage{xcolor}
\usepackage{geometry}
\usepackage{physics}
\usepackage{dcolumn}
\usepackage{bm}
\newcommand{\eqN}[1]{Eq.~(\ref{#1})} 

\begin{document}

\preprint{APS/123-QED}

\title{Nonreciprocal Phase Shifts in Spatiotemporally Modulated Systems}

\author{Jiuda Wu}
\author{Behrooz Yousefzadeh}%
\email{behrooz.yousefzadeh@concordia.ca}
\affiliation{%
Department of Mechanical, Industrial and Aerospace Engineering, Concordia University\\
1455 De Maisonneuve Blvd. W. EV-4.139, Montreal, QC, H3G 1M8, CANADA
}
\date{\today}

\begin{abstract}
Materials and devices subject to spatiotemporal modulation of their effective properties have a demonstrated ability to support nonreciprocal transmission of waves. Most notably, spatiotemporally modulated systems can restrict wave transmission to only one direction; {\it i.e.} a very large difference in the energy transmitted between two points in opposite directions. Taking on a different perspective on nonreciprocity, we here present a response regime in spatiotemporally modulated systems that is characterized by equal transmitted amplitudes (energies) but different phases. The only contributor to nonreciprocity is therefore the nonreciprocal phase shift, the difference between the transmitted phases in the opposite directions. We develop a methodology for realization of nonreciprocal phase shifts based on the response envelopes. This includes a formulation that ensures the same transmitted waveform, along with a special case of near-reciprocal transmission. We focus primarily on steady-state vibration transmission in short, weakly modulated systems, but include a special case of nonreciprocal phase shifts for systems with arbitrary length and strength of modulation. We discuss the main limitations of our methodology, as well as a pathway to overcome it, to motivate further developments on strongly modulated systems. While showcasing a new way for controlling vibration information transmission, our findings highlight the potential role of phase as an additional parameter in nonreciprocal transmission in spatiotemporally modulated systems.  
\end{abstract}

\maketitle


\section{Introduction}
Reciprocity theorems state that wave propagation in a material is independent of the direction of transmission: the transmitted wave remains unchanged if the locations of the source and receiver are interchanged~\cite{Recip_book}. In analysis of the steady-state response, reciprocity manifests as the symmetry of the transfer matrix between the input and output~\cite{newland}. This property has laid the foundation for several measurement techniques and industrial applications~\cite{ewins,flexibility,Recip_app1,Recip_app2,Recip_app3}. Recently, however, there has been a surge of interest in realizing direction-dependent transmission properties in acoustic and mechanical systems, a feat that is impossible within the framework of reciprocity~\cite{NRM2020,bandaru}. 

In linear systems, one way to realize nonreciprocal wave transmission is through {\it spatiotemporal modulations}: periodic changes in the effective properties of the medium in both space and time~\cite{modu_nonrecp_book}. 
To enable nonreciprocity, spatiotemporal modulations are usually introduced to the effective elasticity (stiffness) in various models of waveguides. Examples include a uniform bar with wavelike spatiotemporal modulation in its elastic modulus~\cite{nonrecp_media1,nonrecp_media2,nonrecp_media3} or with local modulated attachments~\cite{nonrecp_media_atch1,nonrecp_media_atch2,nonrecp_media_atch3}, discrete periodic materials with spatiotemporally modulated coupling springs~\cite{nonrecp_mass-spring-chain} or grounding springs~\cite{nonrecp_meta_gd,nonrecp_exp_mag1}, and metamaterials with spatiotemporal modulations in the stiffness of the local resonant springs~\cite{nonrecp_meta_res1,nonrecp_meta_res2} or springs of surface oscillators~\cite{nonrecp_meta_os1,nonrecp_meta_os2}. Nonreciprocal vibration transmission can also occur in systems with spatiotemporal modulations in the effective inertia (masses)~\cite{nonrecp_meta_ms} or in systems with two-phase modulations, {\it i.e.} in both masses and springs~\cite{nonrecp_meta_mssp} or in both density and elastic modulus~\cite{nonrecp_2phase1,nonrecp_2phase2}. By introducing spatiotemporal modulations to the electrical boundary conditions of each cells in piezoelectric media, directional elastic waves can be generated and result in direction-dependent propagation of incident waves~\cite{piezo_croenne2019,piezo_tessier2023}. 

Successful realization of nonreciprocal dynamics leads to the dependence of at least one of the transmission characteristics (amplitude, phase, phase or group velocity, etc.) on the direction of transmission.  The difference between the transmitted amplitudes or energies in the opposite directions is almost universally used as the indicator of nonreciprocity. Typical ways to illustrate this effect are directional bandgaps~\cite{nonrecp_media1,nonrecp_media3,nonrecp_meta_gd,nonrecp_meta_ms,nonrecp_exp6}, frequency spectra of transmitted vibrations~\cite{nonrecp_media_atch1,nonrecp_exp_mag1,nonrecp_meta_res2,nonrecp_meta_os2,nonrecp_exp1_1} or the temporal response of the system~\cite{nonrecp_media1,nonrecp_exp2,nonrecp_exp_mag1}. 

The difference between the transmitted phases can also contribute to nonreciprocity. Phase has played an important role in nonreciprocal propagation of electromagnetic waves~\cite{nonrecp_electr,optics_review}. When the propagation direction is reversed, significant phase difference can be observed in electromagnetic waves that propagate through a time-invariant waveguide inside a magnetic field~\cite{YIG_thinfilm,Magnet_waveguide,phase_reflect,phase_reflect2} or a spatiotemporally modulated waveguide~\cite{phase_nonmagnet}. Several techniques have been developed to realize direction-dependent phases of electromagnetic waves, with current or potential applications in industries such as telecommunications~\cite{nonrecp_electr_tele1,nonrecp_electr_tele2,nonrecp_electr_5G}, radar systems~\cite{nonrecp_electr_radar} and medical magnetic resonance imaging~\cite{nonrecp_electr_MRI}. In acoustic and mechanical systems, however, this attribute of nonreciprocity (contributions of phase) remains unexplored in comparison. 
 
One way to highlight the role of phase in breaking reciprocity is to identify nonreciprocal response regimes that are characterized by equal energies transmitted in opposite directions. In this case, because the transmitted energies are the same, the difference between the phases of the transmitted waves is the only contributor to nonreciprocity. We refer to this scenario as {\it phase nonreciprocity}, and to the corresponding difference between the transmitted phases as the {\it nonreciprocal phase shift}. Phase nonreciprocity has been reported in the steady-state response of nonlinear systems to harmonic excitation~\cite{nonlinear2022,nonlinear2024}, where the nonreciprocal phase shift can be controlled by varying system parameters. In spatiotemporally modulated materials, nonreciprocal phase shifts have been identified as the main contributor to nonreciprocity in systems with a small number of modulated units~\cite{paper1}. A detailed analysis of nonreciprocal phase shifts for modulated systems, however, remains to be presented. 

In this work, our goal is to systematically characterize nonreciprocal phase shifts in discrete, spatiotemporally modulated systems. We focus on the steady-state response of a one-dimensional (1-D) systems subject to simultaneous spatiotemporal modulation and external harmonic drive. Due to the scattering effects of the modulations, the response of a modulated system is not periodic in nature (characterized by two incommensurate frequencies). Therefore, the analysis of the response, and thereby the direction-dependent propagation of phase, is not straightforward and the methodology used for nonlinear systems~\cite{nonlinear2022} is not directly applicable here. We therefore develop a methodology based on the envelopes of the transmitted vibrations in order to enable a systematic search for and realization of nonreciprocal phase shifts in spatiotemporally modulated systems. This includes a near-zero nonreciprocal phase shift and retrieval of reciprocal response. While the primary focus of the work is on weakly modulated systems with a small number of units, we also present a special case of the phenomenon for systems of arbitrary length and strength of modulation. 

Section II provides a derivation of response envelopes for the steady-state response of the system, along with a short overview of nonreciprocity. In Section III, we present our methodology for obtaining response that exhibits nonreciprocal phase shifts, followed by a constraint that ensures the same shapes for the two response envelopes. Section IV provides two special cases of nonreciprocal phase shifts: systems with arbitrary length and strength of modulation, and near-reciprocal transmission. We discuss the limitations of our methodology for obtaining nonreciprocal phase shifts in Section V. Section VI summarizes our findings.

\section{Steady-state Response of the Modulated Systems}

\subsection{Equations of motion}
\begin{figure}[htb]
\centerline{\includegraphics[scale=0.4]{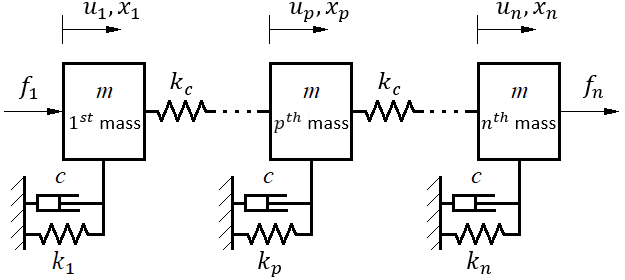}}
\caption{Schematic of the modulated system with $n$ DoF.}
\label{fig_nDoF}
\end{figure}

Fig.~\ref{fig_nDoF} shows the schematic of the discrete model of the spatiotemporally modulated material that we study in this work. The model consists of $n$ identical masses, viscous dampers, coupling springs and modulated grounding springs. The longitudinal rectilinear movement of each mass is considered as its only degree of freedom (DoF). External harmonic forces are applied on the first mass and the last mass, $f_1(t)=F_1\cos{(\omega_f t)}$ and $f_n(t)=F_n\cos{(\omega_f t)}$. The stiffness coefficient of each grounding spring is composed of a constant term and a time-periodic term, expressed as $k_p(t)=k_{g,DC}+k_{g,AC}\cos{(\omega_m t-\phi_p)}$, where $\phi_p=(p-1)\phi$ and $p=1,2,\cdots,n$. Parameter $\omega_m$ is the modulation frequency and $\phi$ is the phase shift between the modulations in two adjacent units. 

Parameter $\phi$ represents the spatial variation of the grounding stiffness along the length of the system. This is the same as the modulation wavenumber in long systems. Nevertheless, we continue referring to $\phi$ as the modulation phase in this work (instead of modulation wavenumber) because we consider systems that can often be too short to even contain one full modulation wavelength within them. The modulation phase is the only parameter that breaks the mirror symmetry of the model and enables nonreciprocity; {\it i.e.}, the end-to-end transmission is always reciprocal for $\phi=0$ by virtue of mirror symmetry.

We start by nondimensionalizing the governing equations, as detailed in Appendix~\ref{appendix:nondimensionalization}. The nondimensional equation of motion for the $p\,$-th mass of the system shown in Fig.~\ref{fig_nDoF} is:
\begin{equation}
\label{eq_EoM_p}
\begin{array}{c}
\ddot{x}_p+2\zeta\dot{x}_p+[1+K_m\cos(\Omega_m\tau-\phi_p)]x_p \vspace{2mm} \\
+K_c \Delta^2_p = P_p\cos(\Omega_f \tau)
\end{array}
\end{equation}
where the overdot represents differentiation with respect to nondimensional time $\tau$. The difference term $\Delta^2_p=2x_p-x_{p-1}-x_{p+1}$ everywhere except at the two ends where $\Delta^2_1=x_1-x_2$ and $\Delta^2_n=x_n-x_{n-1}$. The external forces are only applied at the two ends: $P_p=P$ for $p\in\{1,n\}$ and zero everywhere else. Only one of the two ends is subject to an external force at a time, as explained in Section~\ref{sec:nonreciprocity}. 

\subsection{Response envelopes}
\label{sec:envelopes}
The response of the modulated system is characterized by two independent frequencies $\Omega_f$ and $\Omega_m$. The steady-state displacement is therefore not periodic in time (called {\it quasi-periodic}), except for the rare cases when the ratio of the two frequencies is a rational number. However, the envelope of the response is in fact periodic in time~\cite{IDETC2023}. In this section, we develop an expression for the response envelope to simplify the ensuing analysis and computation. 

The steady-state response of the system can be expressed in general as a combination of harmonic components in complex or real notation as:
\begin{equation}
\label{eq_resp1}
\begin{array}{c}
x_{p}(\tau)=\sum^{\infty}_{q=-\infty} [y_{p;q}e^{i(\Omega_f+q\Omega_m)\tau}+c.c.] \vspace{2mm} \\
=\sum^{\infty}_{q=-\infty} 2|y_{p;q}|\cos{\left[\left(\Omega_f+q\Omega_m\right)\tau+\psi_{p;q}\right]}.
\end{array}
\end{equation}
where $y_{p;q}$ is the complex amplitude of each harmonic component and $c.c.$ represents the corresponding complex conjugate terms. The phase in the real representation, $\psi_{p;q}$, can be expressed as:
\begin{equation}
\label{eq_psi}
\begin{array}{c}
\psi_{p;q}=\atan\!2\left(\Im\left(y_{p;q}\right),\Re\left(y_{p;q}\right)\right),
\end{array}
\end{equation}
where $\Re(\ )$ and $\Im(\ )$ indicate the real part and imaginary parts of the complex amplitudes, respectively. 
For a given $\Omega_f$ and a set of system parameters, the complex amplitudes $y_{p;q}$ can be calculated using the averaging method. This results in a linear system of algebraic equations in the complex amplitudes 
\begin{equation}
    [D] \{Y\}=\{F\}
\end{equation}
where the matrix $[D]$ contains the system parameters, $\{Y\}$ is the vector of unknown complex amplitudes and $\{F\}$ contains information about the location and amplitude of the external force. 
The details of this well established methodology are discussed elsewhere~\cite{Trainiti_HB,NJP2023_Ruzzene,paper1} and are not repeated here. Instead, we directly proceed with obtaining the envelope equations. 

The steady-state displacements can alternatively be rewritten as:
\begin{equation}
\label{eq_resp2}
\begin{array}{c}
x_p(\tau)= E_p(\tau)\  C_p(\tau),
\end{array}
\end{equation}
where $E_p(\tau)$ represents the response envelope and $C_p(\tau)$ represents the corresponding carrier wave of unit amplitude. $E_p(\tau)$ is expressed as:
\begin{equation}
\label{eq_envlp}
\begin{array}{c}
E_p(\tau)=2 \big| \sum^{\infty}_{q=-\infty} y_{p;q} e^{iq\Omega_m\tau} \big|.
\end{array}
\end{equation}
When $\Omega_m<\Omega_f$ and $K_m\leq0.1$, $C_p(\tau)$ can be approximated by a harmonic wave of the same frequency as the external excitation:
\begin{equation}
\label{eq_RespCarr}
\begin{array}{c}
C_p(\tau)= \cos{(\Omega_f\tau+\psi_{p;0})},
\end{array}
\end{equation}
where $\psi_{p;0}=\atan\!2\left(\Im\left(y_{p;0}\right),\Re\left(y_{p;0}\right)\right)$. Otherwise, $C_p(\tau)$ is not periodic but its amplitude remains equal to $1$. 

We emphasize that although the steady-state displacements $x_p(\tau)$ are not periodic in time, their envelopes, $E_p(\tau)$, remain periodic with period $T_E=2\pi/\Omega_m$. Regardless, the displacement function $x_p(\tau)$ remains bounded by its envelopes $\pm \,E_p(\tau)$. 

The infinite summations in Eqs.~(\ref{eq_resp1}) and~(\ref{eq_envlp}) need to be truncated at a finite value of $q$. We take an expansion with $q\in[-\mathcal{F},\mathcal{F}]$ and $\mathcal{F}\in\mathbf{N}$. We restrict our attention to short systems with weak modulations, {\it i.e.} $n\leq5$ and $K_m\leq0.1$. Under these assumptions, the response envelope is nearly harmonic. Thus, we use $\mathcal{F}=1$ to approximate the steady-state response of the system. 

\begin{figure}[htb]
\centerline{\includegraphics[scale=0.35]{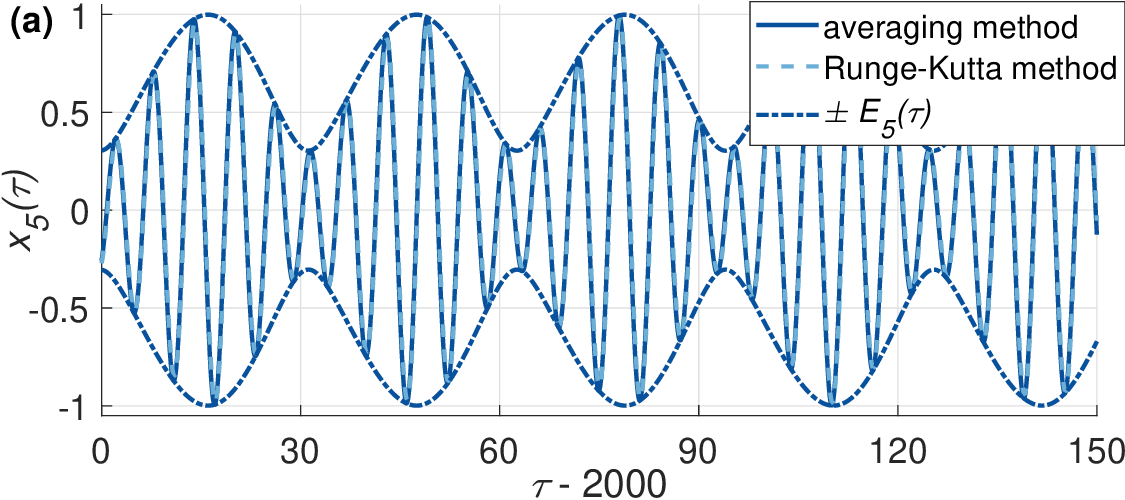}}
\centerline{\includegraphics[scale=0.35]{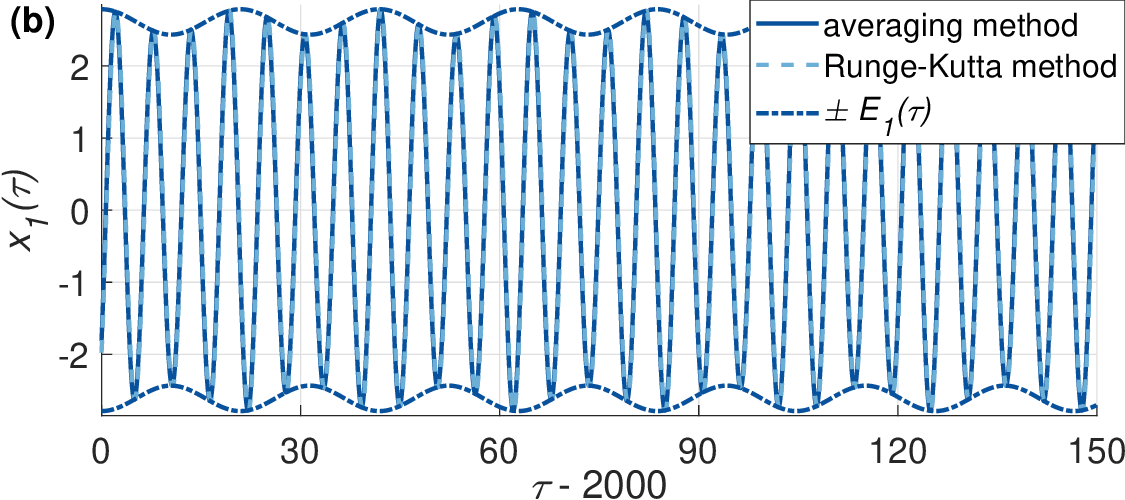}}
\caption{Comparison between the output displacements computed using the averaging method (solid curves) and the Runge-Kutta method (dashed curves). Dash-dotted curves are plots of $\pm \,E_5(\tau)$ and $\pm \,E_1(\tau)$ in panels (a) and (b), respectively.
(a) $n=5$, $\Omega_m=0.2$, $\phi=0.42\pi$, $\Omega_f=0.88$, $K_c=0.6$, $K_m=0.1$, $\zeta=0.02$, $P_1=1$ and $P_5=0$; (b) $n=4$, $\Omega_m=0.3$, $\phi=0.95\pi$, $\Omega_f=1.1$, $K_c=0.8$, $K_m=0.1$, $\zeta=0.02$, $P_1=0$ and $P_4=1$.}
\label{fig_comp}
\end{figure}

To validate the solution predicted by Eq.~\eqref{eq_resp1} with $q\in[-1,1]$, the response of Eq.~\eqref{eq_EoM_p} is computed using the Runge-Kutta method until the steady state is reached. We arbitrarily choose two sets of system parameters, then calculate the displacements of two masses using both the averaging and Runge-Kutta methods, as shown in Fig.~\ref{fig_comp}. We observe that the averaging method predicts the steady-state response well, $\pm \,E_5(\tau)$ and $\pm \,E_1(\tau)$ following the response envelopes of the outputs accurately in both cases. 

The response envelope for a system with stronger modulation amplitude remains periodic but it is no longer harmonic~\cite{IDETC2023}. The formulation in Eqs.~\eqref{eq_envlp} can still accurately predict the response envelopes provided that more terms are included in the expansion ($\mathcal{F}>1$). In contrast, more conventional methods of computing the response envelopes based on the method of multiple scales or rotating wave approximation are typically limited to harmonic envelopes~\cite{SRWA}. 

\subsection{Nonreciprocity} 
\label{sec:nonreciprocity}
Before presenting the special case of nonreciprocal phase shifts, we briefly review the typical scenario for nonreciprocal vibration transmission in our system. 
The response of an externally forced system is reciprocal if it remains invariant upon interchanging the locations of the input (source) and output (receiver). 
To test for reciprocity, therefore, we define two configurations to distinguish between the two directions of vibration transmission: (i) the {\it forward} (from left to right) configuration with $P_1=P$ and $P_n=0$, where the output is the steady-state displacement of the rightmost (last) mass, $x_n^F(\tau)$; (ii) the {\it backward} (from right to left) configuration with $P_1=0$ and $P_n=P$, where the output is the steady-state displacement of the leftmost (first) mass, $x_1^B(\tau)$. The superscripts $F$ and $B$ denote the response in the {\it forward} and {\it backward} configurations, respectively. Vibration transmission is then reciprocal if and only if $x_n^F(\tau)=x_1^B(\tau)$. 

We introduce the reciprocity bias, $R$, to quantify the degree of nonreciprocity of the response:
\begin{equation}
\label{eq_R}
\begin{array}{c}
\displaystyle R = \lim_{T\to\infty} \sqrt{ \frac{\int_{0}^{T} \left[x_n^F\left(\tau\right) - x_1^B\left(\tau\right)\right]^2 \, \dd{\tau}}{2\int_{0}^{T} \left[x_n^F\left(\tau\right) \right]^2 + \left[x_1^B\left(\tau\right)\right]^2 \, \dd{\tau}} } \vspace{2mm} \\
\displaystyle = \sqrt{ \frac{\sum^{\mathcal{F}}_{q=-\mathcal{F}} \left| y_{n;q}^F - y_{1;q}^B \right|^2}{2\sum^{\mathcal{F}}_{q=-\mathcal{F}} \left| y_{n;q}^F\right|^2 + \left|y_{1;q}^B \right|^2} }.
\end{array}
\end{equation}
By definition, $0 \leq R \leq 1$, and $R=0$ if and only if vibration transmission through the system is reciprocal. The case of $R=1$ corresponds to unilateral transmission, which is not relevant in this work. The denominator in Eq.~(\ref{eq_R}) is introduced to remove the apparent increase in the degree of nonreciprocity due to $P$. Because we study a linear system in this work, the reported value of $R$ hold for any choice of $P$. 

\begin{figure}[htb]
\centerline{\includegraphics[scale=0.35]{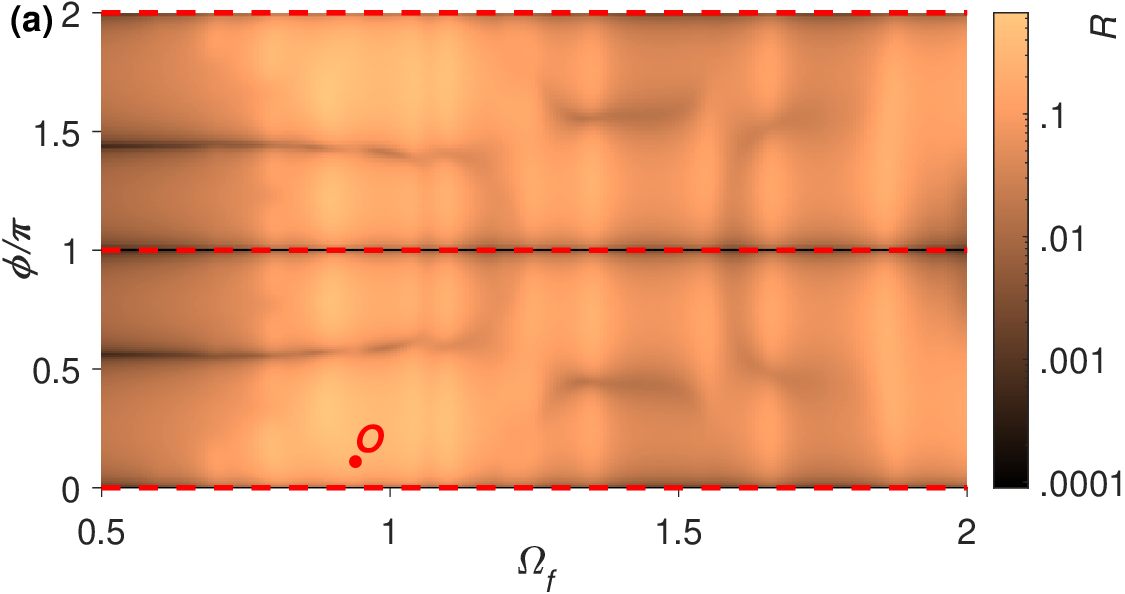}}
\centerline{\includegraphics[scale=0.35]{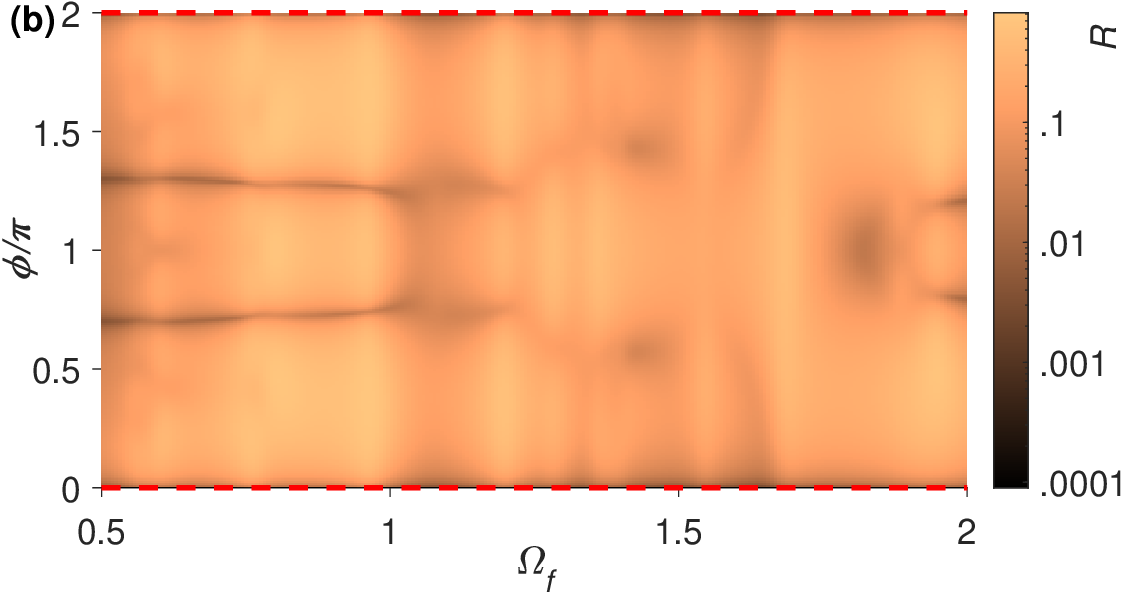}}
\caption{Surface plots of reciprocity bias as a function of $\Omega_f$ and $\phi$ for $K_m=0.1$, $\zeta=0.02$ and $P=1$. Red dashed lines indicate the combinations of $\Omega_f$ and $\phi$ that lead to $R=0$. (a) $n=5$, $\Omega_m=0.1$ and $K_c=0.8$; (b) $n=4$, $\Omega_m=0.2$ and $K_c=0.6$.}
\label{fig_R}
\end{figure}

To demonstrate the influence of $\phi$ on nonreciprocity, we compute $R$ as a function of $\phi$ and $\Omega_f$ for two sets of system parameters; see Fig.~\ref{fig_R}. 
In general, the response remains nonreciprocal ($R>0$) over the entire range $0<\phi<2\pi$. A prominent exception is for systems with an odd number of units, for which the response is reciprocal ($R=0$) for $\phi=\pi$; see Fig.~\ref{fig_R}(a). This is because the time-reversal symmetry is retained when $\phi=\pi$ and $n$ is an odd number. The role of phase on nonreciprocity in coupled spatiotemporally modulated systems is explored in detail elsewhere~\cite{paper1}. 
\begin{figure}[htb]
\centerline{\includegraphics[scale=0.35]{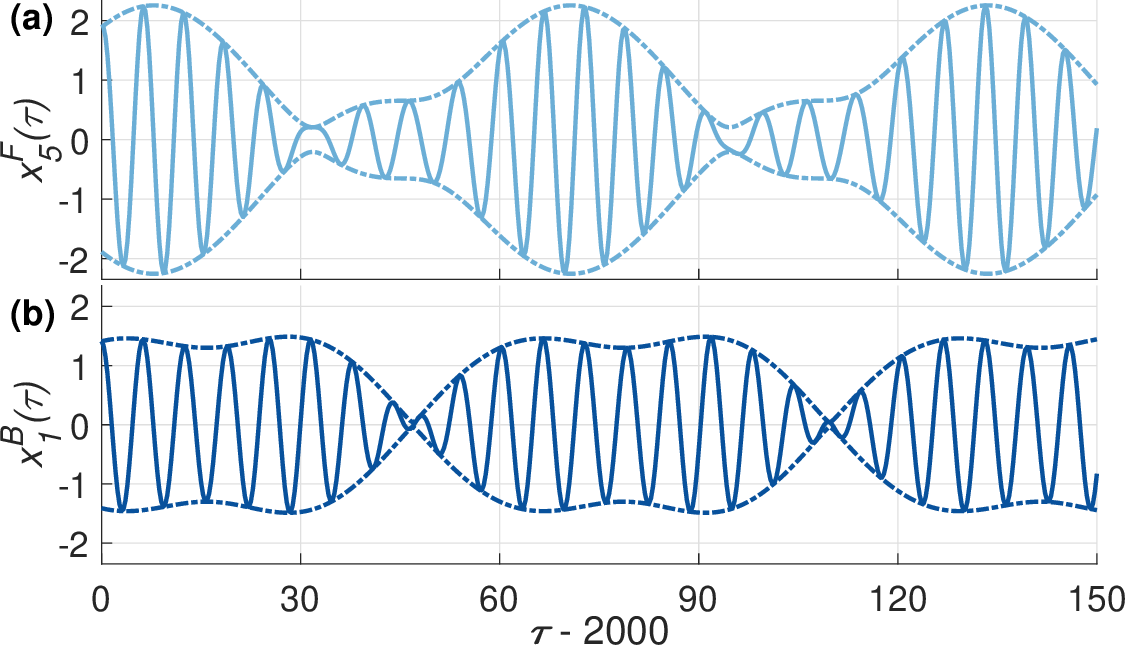}}
\centerline{\includegraphics[scale=0.35]{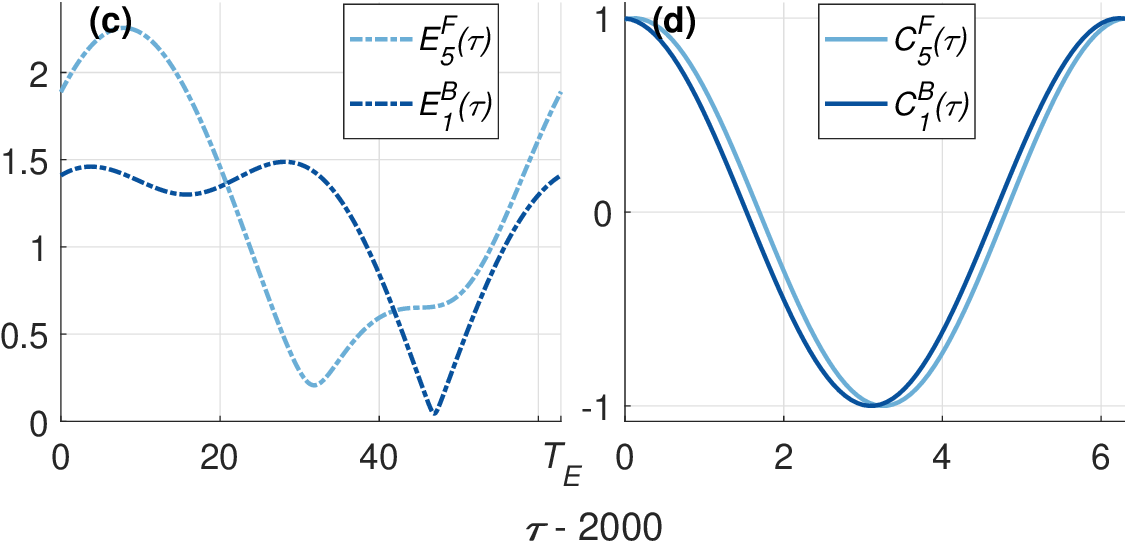}}
\caption{Plots of (a,b) output displacements, (c) response envelopes and (d) carrier waves for point $O$. Dash-dotted curves are plots of $\pm \,E_5^F(\tau)$ and $\pm \,E_1^B(\tau)$ in panels (a) and (b), respectively.
System parameters: $n=5$, $\Omega_m=0.1$, $\phi=0.11\pi$, $\Omega_f=0.94$, $K_c=0.8$, $K_m=0.1$, $\zeta=0.02$ and $P=1$.}
\label{fig_R-example}
\end{figure}

Fig.~\ref{fig_R-example} shows the response of the system at point $O$ of Fig.~\ref{fig_R}(a) with $\Omega_f=0.94$ and $\phi=0.11\pi$. Vibration transmission is clearly nonreciprocal, as seen in panels~(a) and~(b). The difference in the transmitted amplitudes is also evident in the response envelopes in panel~(c). Panel~(d) shows that, in addition to the response envelopes, the two carrier waves are also different between the two configurations, thus contributing to nonreciprocity. The carrier waves are harmonic in weakly modulated systems; recall Eq.~(\ref{eq_RespCarr}). 

The primary focus of this work is to investigate nonreciprocal response with equal amplitudes transmitted in the opposite directions. Because the amplitudes are the same, nonreciprocity is caused merely by the phase difference between the transmitted vibrations: nonreciprocal phase shift. In time-invariant nonlinear systems, it is possible to realize nonreciprocal phase shifts in the steady-state response to external harmonic excitation~\cite{nonlinear2022,nonlinear2024}. Nonreciprocal phase shifts occur when the two outputs have the same amplitudes but different phases, where the amount of the nonreciprocal phase shift can be controlled using a system parameter. In spatiotemporally modulated systems, however, a similarly systematic investigation of nonreciprocal phase shifts is more challenging due to the non-periodic nature of the response. We will take advantage of the periodicity of the response envelopes to obtain a vibration transmission scenario that exhibits nonreciprocal phase shifts. 

\section{Nonreciprocal Phase Shifts}
\label{sec:nonrecpPhase}
Because the steady-state response of an externally driven modulated system is quasi-periodic, the definition of its phase is not as straightforward as in systems without modulation. This further complicates the search for parameters that lead to nonreciprocal phase shifts in the response of the system. Given that the envelope of the response remains periodic, it is much more feasible to investigate nonreciprocal phase shifts based on the response envelopes. We expect that the steady-state displacements exhibit a nonreciprocal phase shift if the corresponding response envelopes exhibit a nonreciprocal phase shift. We discuss this approach here. 

\subsection{Formulation}
The displacement response of the system oscillates in time with two incommensurate frequencies around the static equilibrium point; {\it i.e.}, there is no bias (DC shift with respect to $x_p=0$) in the response. As a result, the envelope equations appear as a pair located symmetrically with respect to $x_p=0$; see the envelopes in Fig.~\ref{fig_comp}. We can therefore express the envelope equation as
\begin{equation}
\label{eq_envlp2}
\begin{array}{c}
E_p(\tau)=\sqrt{\mathcal{S}_{dc;p}+\mathcal{S}_{ac;p}(\tau)},
\end{array}
\end{equation}
where $\mathcal{S}_{dc;p}$ represent time-independent bias portions (DC shifts) and $\mathcal{S}_{ac;p}$ represent the time-varying portions of the amplitude equation. The square root is included in Eq.~(\ref{eq_envlp2}) because we will be working with the square of envelope, $E_p^2(\tau)$. With $\mathcal{F}=1$, the envelope bias terms are defined as:
\begin{equation}
\label{eq_S_dc}
\begin{array}{c}
\displaystyle \mathcal{S}_{dc;p} = \frac{1}{T_E} \int_{0}^{T_E} \left(E_p\left(\tau\right)\right)^2 \, \dd{\tau} \vspace{2mm} \\
\displaystyle = 4\left(\big|y_{p;-1}\big|^2+\big|y_{p;0}\big|^2+\big|y_{p;1}\big|^2\right).
\end{array}
\end{equation}
And the time-varying portion of the envelopes are defined as:
\begin{equation}
\label{eq_S_ac}
\begin{array}{c}
\mathcal{S}_{ac;p}(\tau) = \mathcal{A}_p\cos{(\Omega_m\tau-\theta_{a;p})} \vspace{2mm} \\
\displaystyle + \mathcal{B}_p\cos{(2\Omega_m\tau-\theta_{b;p})},
\end{array}
\end{equation}
See Appendix~\ref{appendix:envelopeAB} for the expressions for the envelope amplitudes, $\mathcal{A}_p$ and $\mathcal{B}_p$, and envelope phases, $\theta_{a;p}$ and $\theta_{b;p}$, in terms of the response amplitudes. 

In terms of the envelope parameters $\mathcal{S}_{dc;n,1}^{F,B}$, $\mathcal{A}_{n,1}^{F,B}$ and $\mathcal{B}_{n,1}^{F,B}$, nonreciprocal phase shift between response envelopes is characterized by the following three constraints:
\begin{subequations}
\label{eq_PhaseNon}
\begin{equation}
\label{eq_PhaseNon_a}
\mathcal{S}_{dc;n}^{F} = \mathcal{S}_{dc;1}^{B}, \vspace{-2.5mm}
\end{equation}
\begin{equation}
\label{eq_PhaseNon_b}
\mathcal{A}_n^F = \mathcal{A}_1^B, \vspace{-1.5mm}
\end{equation}
\begin{equation}
\label{eq_PhaseNon_c}
\mathcal{B}_n^F = \mathcal{B}_1^B.
\end{equation}
\end{subequations}
If the three constraints in Eqs.~\eqref{eq_PhaseNon} are satisfied, it is still possible that the two envelopes are not the same, $E_n^F(\tau) \neq E_1^B(\tau)$. This can happen because of the phase terms in Eq.~\eqref{eq_S_ac}; {\it i.e.}, $\theta_{a,n}^F \neq \theta_{a,1}^B$ or $\theta_{b,n}^F \neq \theta_{b,1}^B$. The envelopes $E_n^F(\tau)$ and $E_1^B(\tau)$ would exhibit a nonreciprocal phase shift in this case, which corresponds to a nonreciprocal phase shift in the displacement as well. Therefore, we use Eqs.~\eqref{eq_PhaseNon} as the constraints that need to be satisfied for nonreciprocal phase shifts. 

We note that satisfying Eqs.~\eqref{eq_PhaseNon} provides a necessary but insufficient condition for the occurrence of nonreciprocal phase shifts. 
Having equal {\it forward} and {\it backward} response envelopes, $E_n^F(\tau) = E_1^B(\tau)$, may not necessarily be equivalent to $x_n^F(\tau) = x_1^B(\tau)$ due to the possible phase difference between two carrier waves. We have not encountered this particular scenario in our simulations, however. 
See Section~\ref{sec:limit2} for the extension of the formulation to $\mathcal{F}\ge1$. 

\subsection{Solution methodology and results}
\label{sec:method_PhaseNon}
To find system parameters that lead to nonreciprocal phase shifts, we first need to calculate the steady-state response amplitudes of the system, $y_{p;q}$. Truncating the expansion in Eq.~\eqref{eq_EoM_p} at $\mathcal{F}$ sideband frequencies, $q\in[-\mathcal{F},\mathcal{F}]$, there will be $2\mathcal{F}+1$ amplitudes for each degree of freedom. For a system with $n$ masses, considering both the {\it forward} and {\it backward} configurations, the solution process involves calculating $2n(2\mathcal{F}+1)$ complex amplitudes. The averaging method outlined in Section~\ref{sec:envelopes} provides the required linear system of $2n(2\mathcal{F}+1)$ algebraic equations in the complex amplitudes. 

In addition, the three constraints in Eqs.~\eqref{eq_PhaseNon} need to be satisfied in order to ensure the envelope equations exhibit nonreciprocal phase shifts. This requires three of the system parameters to vary independently (free parameters). Not all the system parameters are suitable for this purpose, however. For example, $n$ does not change continuously, $P$ scales the amplitudes linearly and has no effect on the nature of the response (because of linearity), $K_m$ is limited to the range of weak modulations ($K_m\leq0.1$) to ensure $\mathcal{F}=1$ provides sufficient accuracy, and $\zeta$ is limited to small values to ensure light damping. We choose $\phi$, $\Omega_f$ and $K_c$ as the free parameters. The modulation frequency, $\Omega_m$, has a smaller permissible range of variation in comparison and is therefore left unchanged in this section -- we use $\Omega_m$ in Section~\ref{sec:sameshape} to control the shape of the envelopes after nonreciprocal phase shift is achieved. 

\begin{figure*}[htb]
\centerline{\includegraphics[scale=0.35]{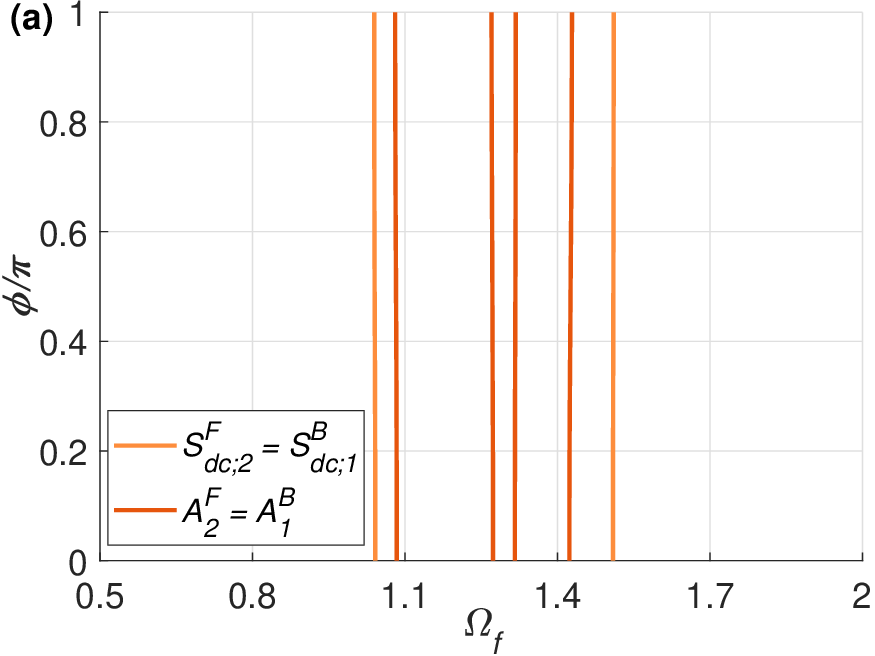}\hspace{1mm}\includegraphics[scale=0.35]{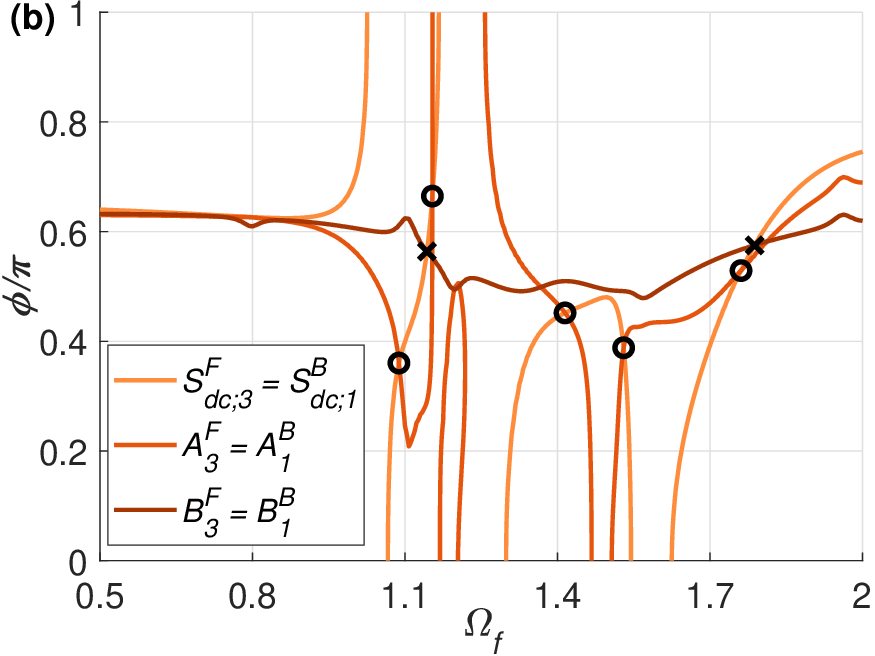}\hspace{1mm}\includegraphics[scale=0.35]{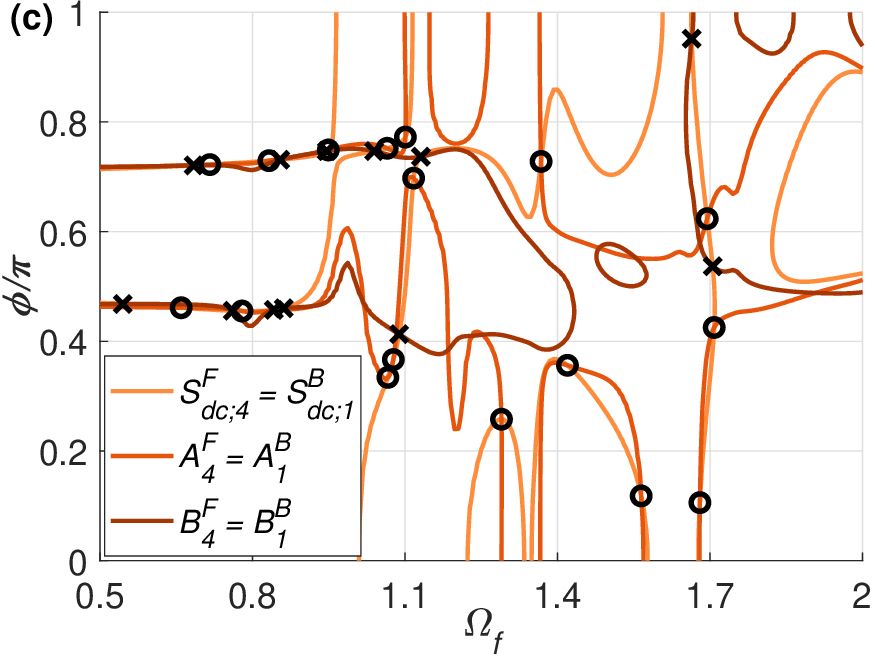}}
\caption{The curves in each panel show the locus of system parameters that satisfy one of the three constraints in Eqs.~\eqref{eq_PhaseNon}. (a) $n=2$, (b) $n=3$ and (c) $n=4$.}
\label{fig_curve}
\end{figure*}
\begin{figure*}[htb]
\centerline{\includegraphics[scale=0.35]{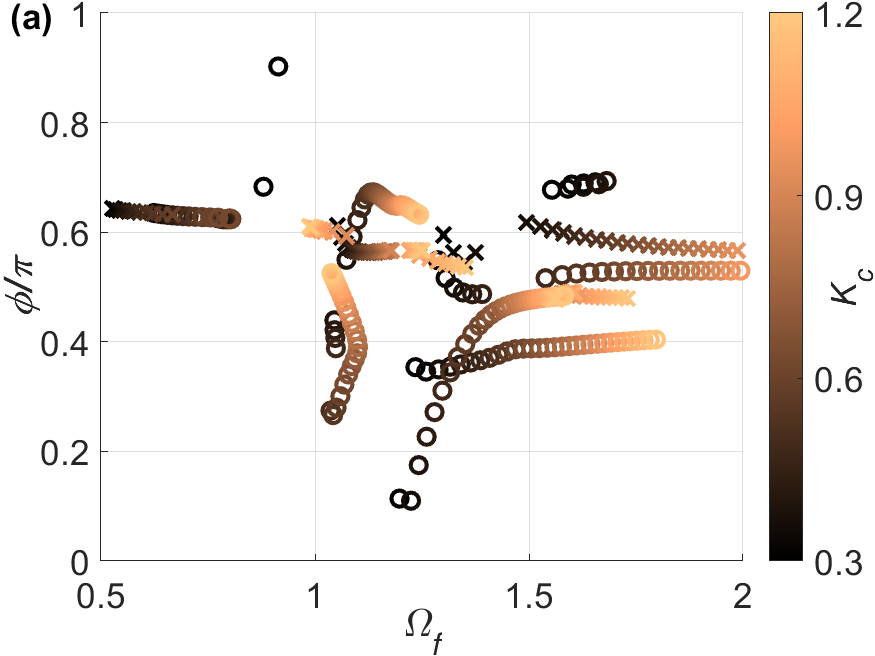}\hspace{1mm}\includegraphics[scale=0.35]{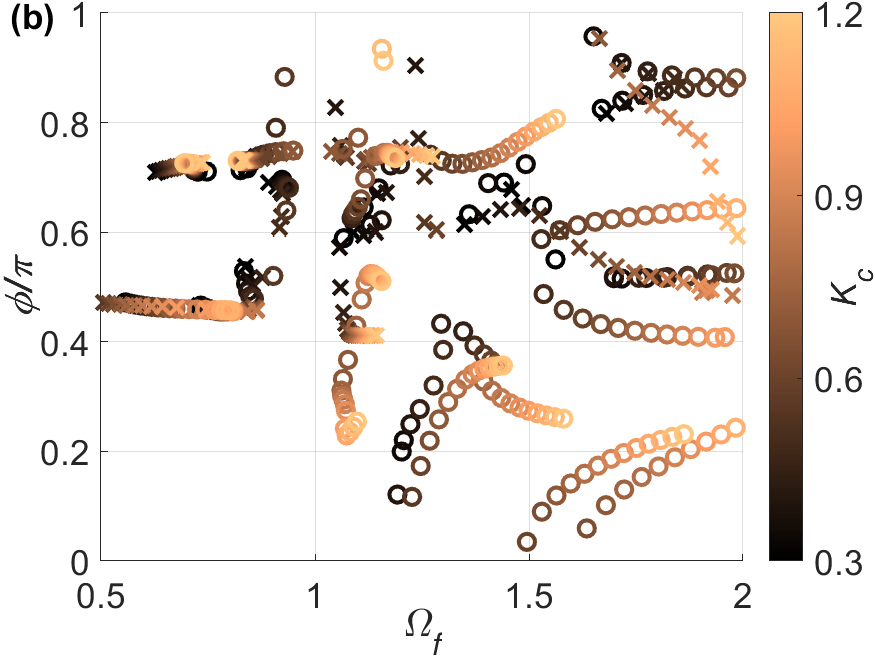}\hspace{1mm}\includegraphics[scale=0.35]{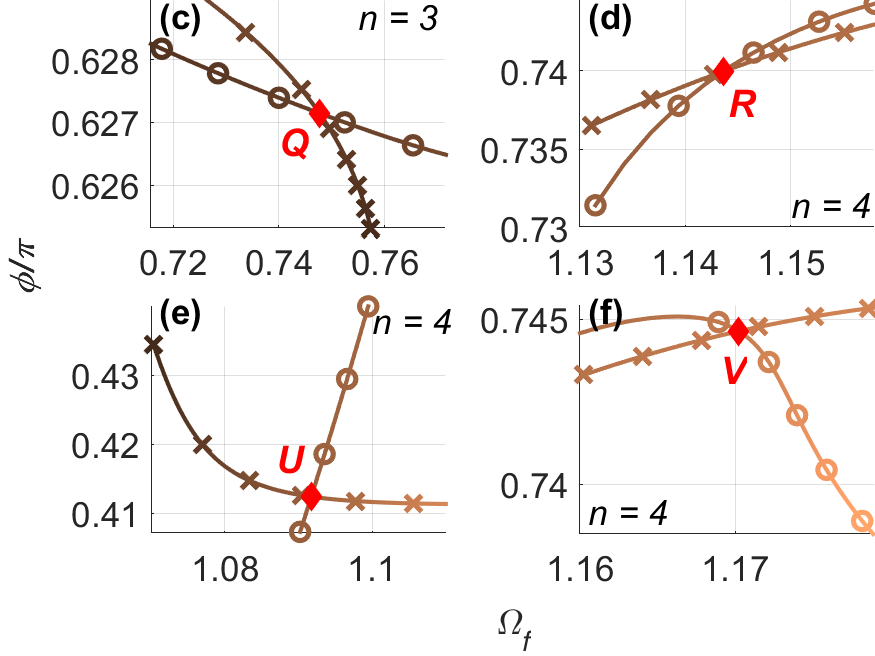}}
\caption{Locus of the intersection points from Fig.~\ref{fig_curve} as a function of $K_c$. (a,c) $n=3$, (b,d-f) $n=4$. Panels (c-f) are zoomed-in views of the intersections points between loci of circles and crosses. The color along each locus indicates the corresponding value of $K_c$, using the same color scale as in (a,b). The intersection points indicated by red diamonds satisfy Eqs.~\eqref{eq_PhaseNon}.}
\label{fig_3Dpoint}
\end{figure*}

This procedure results in a system of $6n$ nonlinear algebraic equations for $\mathcal{F}=1$. Solving this system of equations results in sets of the free system parameters $\phi$, $\Omega_f$ and $K_c$ that lead to nonreciprocal phase shifts. To demonstrate the methodology, we fix the other system parameter to $\Omega_m=0.2$, $K_m=0.1$, $\zeta=0.02$ and $P=1$. We focus on short systems with $n\in\{2,3,4\}$ in this section -- we discuss longer systems in Sections~\ref{sec:special_1} and~\ref{sec:limit2}. 

To satisfy the three constraints in Eqs.~(\ref{eq_PhaseNon}), we first fix $K_c=0.7$ and perform an exhaustive search to find combinations of $\phi$ and $\Omega_f$ that satisfy each pair of the constraint equations over the range $0<\phi<\pi$ and $0.5<\Omega_f<2$. Because of the symmetries of trigonometric functions, the interval from 0 to $\pi$ covers the entire range of variation for $\phi$. We also note that the interval chosen for the forcing frequency covers the primary resonances and sidebands; the frequency range beyond this interval is off-resonance and therefore of limited practical interest. 

Fig.~\ref{fig_curve} shows the combinations of $\phi$ and $\Omega_f$ that satisfy different constraints in Eqs.~\eqref{eq_PhaseNon}. For the system with $n=2$, panel~(a), there are no intersections between the loci of $\mathcal{S}_{dc;2}^F=\mathcal{S}_{dc;1}^B$ and $\mathcal{A}_2^F=\mathcal{A}_1^B$, and no parameter values that result in $\mathcal{B}_2^F=\mathcal{B}_1^B$. Thus, a system with $n=2$ cannot exhibit nonreciprocal phase shifts. 

In systems with more modulated units, the loci of the constraint equations are more complex and several intersection points are possible. Figs.~\ref{fig_curve}(b)~and~\ref{fig_curve}(c) show these intersection points for systems with $n=3$ and $n=4$, respectively. The circle markers indicate intersections points that satisfy both Eq.~\eqref{eq_PhaseNon_a} and Eq.~\eqref{eq_PhaseNon_b}, and cross markers indicate intersection points that satisfy both Eq.~\eqref{eq_PhaseNon_a} and Eq.~\eqref{eq_PhaseNon_c}. 

\begin{figure*}[htb]
\centerline{\includegraphics[scale=0.35]{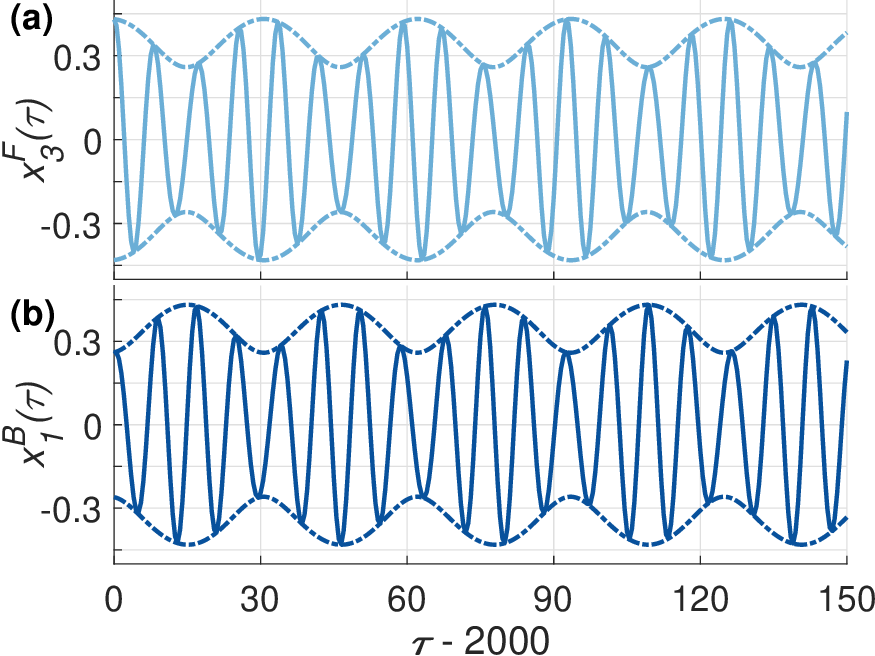}\hspace{1mm}\includegraphics[scale=0.35]{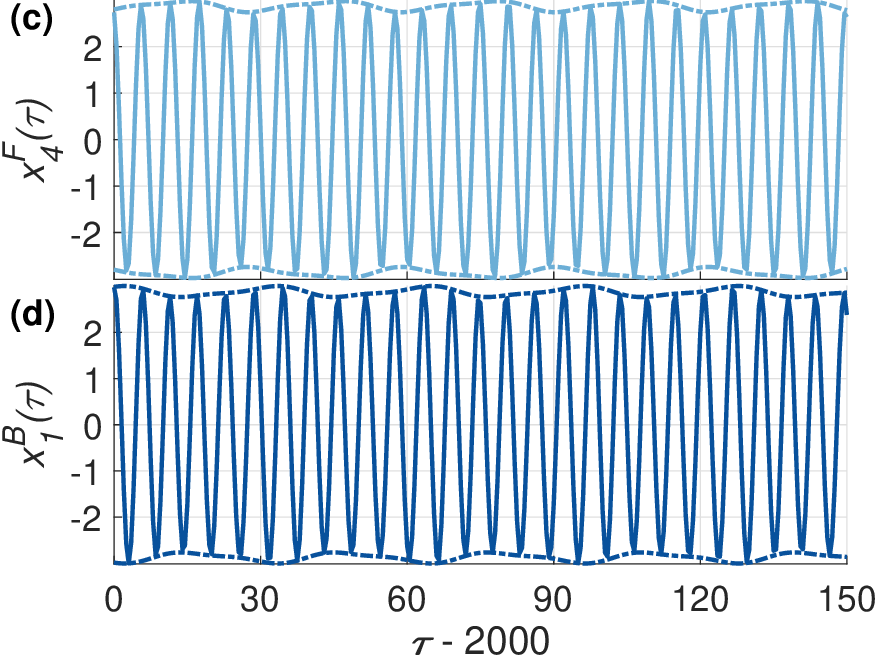}\hspace{1mm}\includegraphics[scale=0.35]{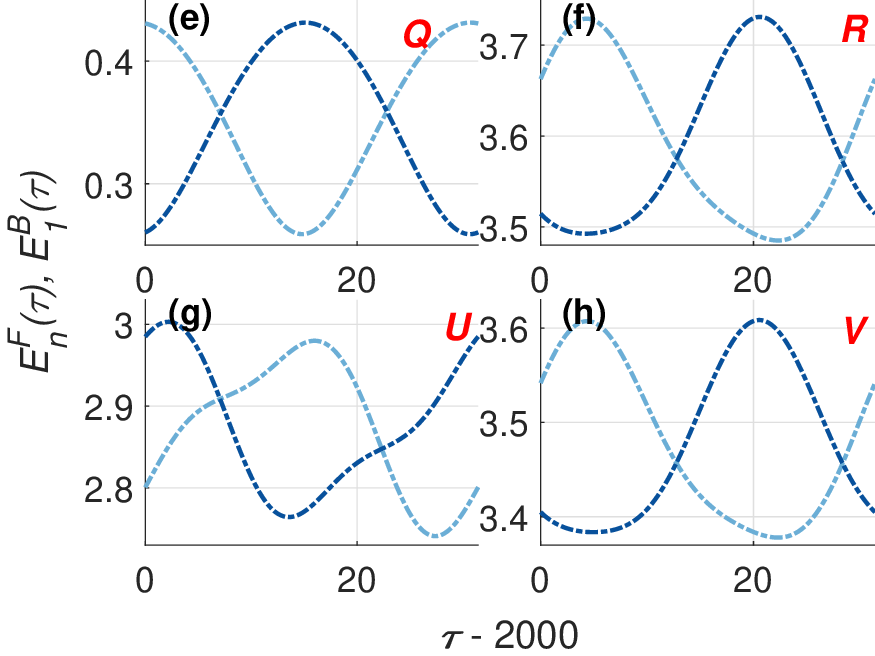}}
\caption{Displacement outputs and response envelopes exhibiting nonreciprocal phase shift for (a,b) point $Q$, (c,d) point $U$. (e-h) Response envelopes for points $Q$, $R$, $U$ and $V$, respectively.}
\label{fig_output}
\end{figure*}

Having found pairs of ($\Omega_f$,$\phi$) that satisfy two of the three constraints (circles and crosses in Fig.~\ref{fig_curve}), we allow $K_c$ to vary and track how the intersection points (circles and crosses) evolve. Figs.~\ref{fig_3Dpoint}(a,b) show the results of this search for systems with $n=3$ and $n=4$, respectively. For better clarity, we have used a colormap in $K_c$ instead of showing the three-dimensional plots. There are several intersection points between the loci of circles and crosses. Figs.~\ref{fig_3Dpoint}(c-f) show closeup views of four of these intersection points, labeled $Q$, $R$, $U$ and $V$. 
These points represent parameters that satisfy the three constraints in Eqs.~\eqref{eq_PhaseNon} and lead to nonreciprocal phase shifts. 

Figs.~\ref{fig_output}(a-d) show the output displacements and the corresponding response envelopes in the time domain for intersection points $Q$ and $U$. As expected, the nonreciprocal phase shift in the response envelopes corresponds to nonreciprocal phase shifts in the displacement outputs. Figs.~\ref{fig_output}(e-h) show the response envelopes, $E_n^F(\tau)$ and $E_1^B(\tau)$, for all the four intersections points identified in Fig.~\ref{fig_3Dpoint}(c-f). None of the four response envelopes in these examples is harmonic, a feature that is particularly visible for point U, Fig.~\ref{fig_output}(g). These points were chosen because they result in significant phase difference between the {\it forward} and {\it backward} configurations. The nonrecipocal phase shift can sometimes be small for other intersection points. 

\subsection{Enforcing the same envelope shapes}
\label{sec:sameshape}
\begin{figure}[b]
\centerline{\includegraphics[scale=0.35]{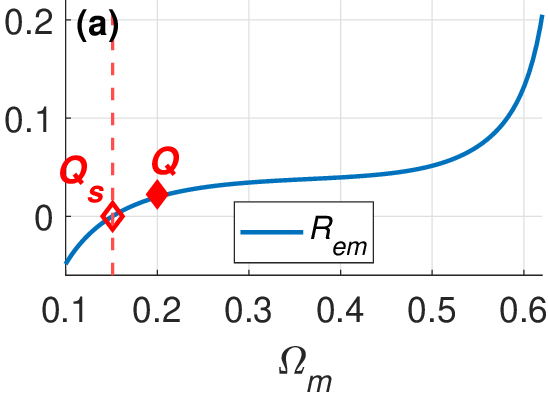}\hspace{1mm}\includegraphics[scale=0.35]{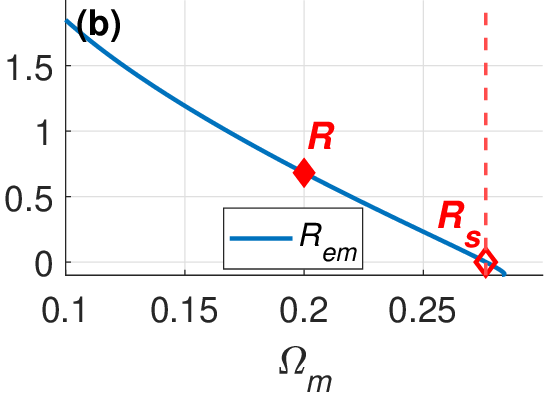}}
\centerline{\includegraphics[scale=0.35]{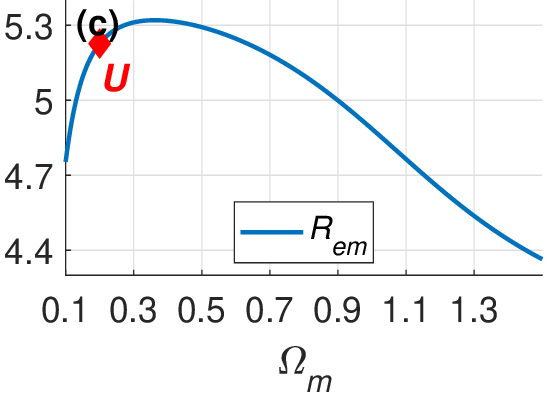}\hspace{1mm}\includegraphics[scale=0.35]{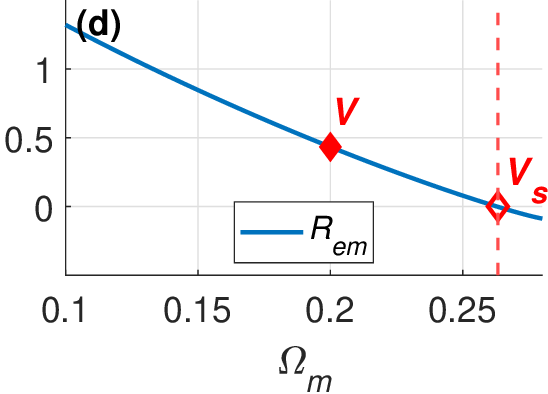}}
\caption{Variation of $\mathcal{R}_{em}$ as a function of $\Omega_m$ for points $Q$ (a), $R$ (b), $U$ (c) and $V$ (d), all exhibiting nonreciprocal phase shifts. $Q_s$, $R_s$ and $V_s$ represent points at which $E_n^F(\tau)$ and $E_1^B(\tau)$ have the same shape.}
\label{fig_WmRLS}
\end{figure}
\begin{figure*}[htb]
\centerline{\includegraphics[scale=0.35]{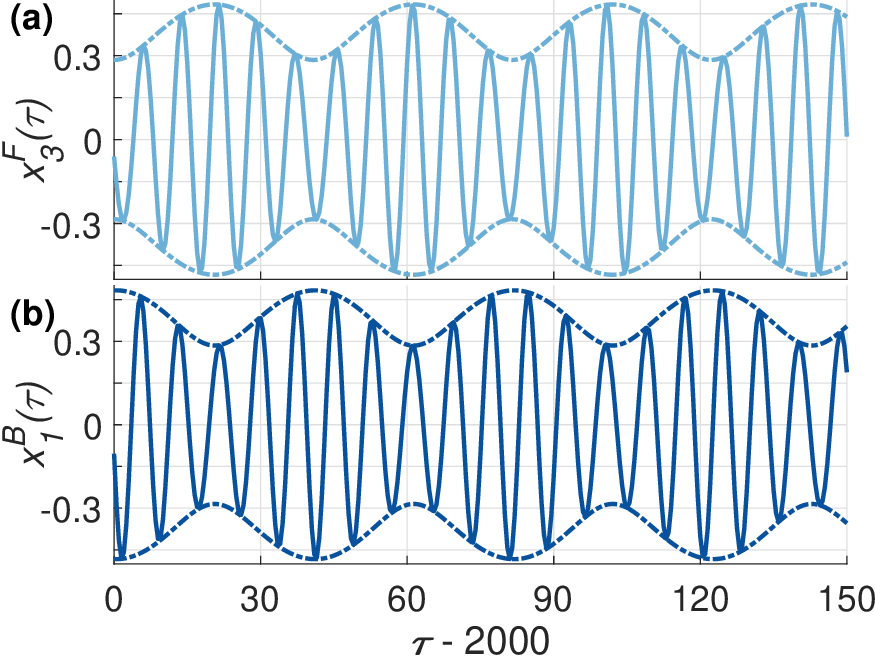}\hspace{1mm}\includegraphics[scale=0.35]{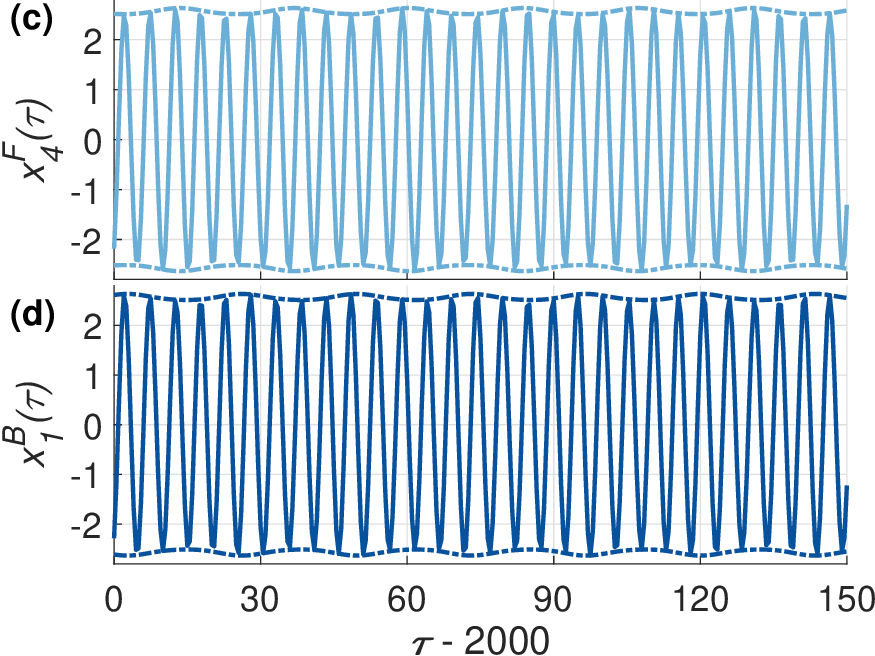}\hspace{1mm}\includegraphics[scale=0.35]{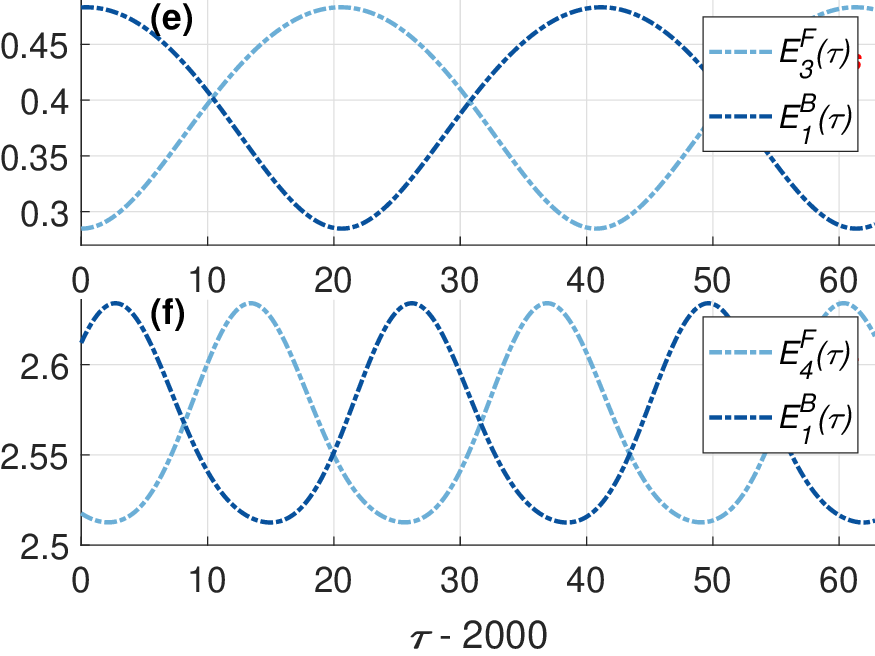}}
\caption{Displacement outputs for points $Q_s$ (a,b) and $V_s$ (c,d) that exhibit nonreciprocal phase shift with the same envelope shapes. The corresponding response envelopes are shown in (e) and (f) respectively.}
\label{fig_output2}
\end{figure*}
The constraints for nonreciprocal phase shift, Eqs.~(\ref{eq_PhaseNon}), do not impose any restrictions on the shape of the response envelopes. The corresponding response envelopes can therefore satisfy the constraints while having different shapes; this is particularly obvious for point $U$, shown in Fig.~\ref{fig_output}(g).
To enforce the same shape for the {\it forward} and {\it backward} response envelopes, we can introduce the following constraint:
\begin{equation}
\label{eq_Eshape}
\begin{array}{l}
\mathcal{R}_{em}=\vspace{1mm} \\
\text{rem}\Bigl(2\bigl(\theta_{a;n}^F-\theta_{a;1}^B\bigr)-\bigl(\theta_{b;n}^F-\theta_{b;1}^B\bigr)+2\hat{z}\pi,2\pi\Bigr)\vspace{1mm} \\
=0,
\end{array}
\end{equation}
where $\text{rem}\left(\alpha,\beta\right)$ is the remainder of dividing $\alpha$ by $\beta$, and $\hat{z}$ is an integer introduced to ensure the continuity of $\mathcal{R}_{em}$. 

To satisfy the additional constraint in Eq.~(\ref{eq_Eshape}), we allow the modulation frequency $\Omega_m$ to vary as a free parameter and monitor $\mathcal{R}_{em}$ for zero crossings. The other system parameters need to vary during this computation to ensure the constraints in Eqs.~(\ref{eq_PhaseNon}) remain satisfied. We note that it may not necessarily be possible to satisfy the additional constraint in Eq.~(\ref{eq_Eshape}) within a reasonable range of $\Omega_m$. 

Fig.~\ref{fig_WmRLS} shows the variation of $\mathcal{R}_{em}$ as a function of $\Omega_m$ for points $Q$, $R$, $U$ and $V$. A set of system parameters exist for which $\mathcal{R}_{em}=0$ for points $Q$, $R$ and $V$, but not for point $U$. We have therefore identified three points, named $Q_s$, $R_s$ and $V_s$, that satisfy all the constraints in Eqs.~(\ref{eq_envlp}) and (\ref{eq_Eshape}). 
Fig.~\ref{fig_output2} shows the displacement outputs and the corresponding response envelopes for points $Q_s$ and $V_s$. As expected, the response envelopes have the same shape and exhibit nonreciprocal phase shift in both cases. 

The results presented in this section were obtained for a fixed set of parameters $n$, $K_m$ and $\zeta$. It is, of course, possible to use the same methodology to find many other sets of parameters at which the response of the system exhibits nonreciprocal phase shifts, potentially with the same envelope shape. 

\section{Special Cases}
\label{sec:specials}

\subsection{Systems with $\phi=\pi$ and even $n$}
\label{sec:special_1}

\begin{figure}[b]
\centerline{\includegraphics[scale=0.35]{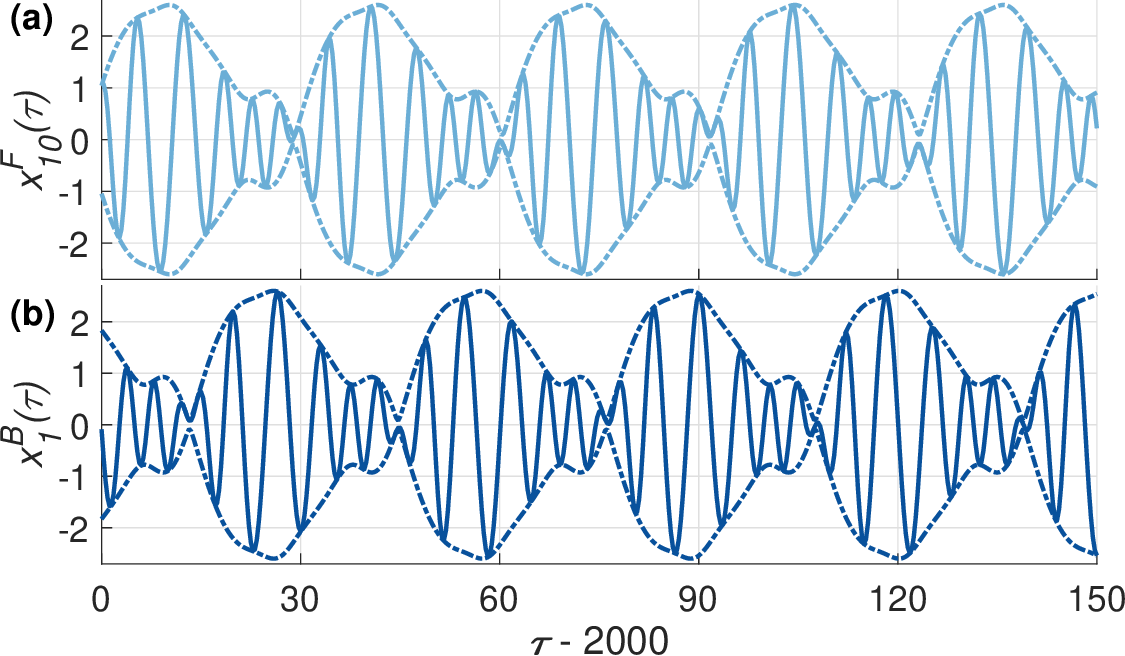}}
\centerline{\includegraphics[scale=0.35]{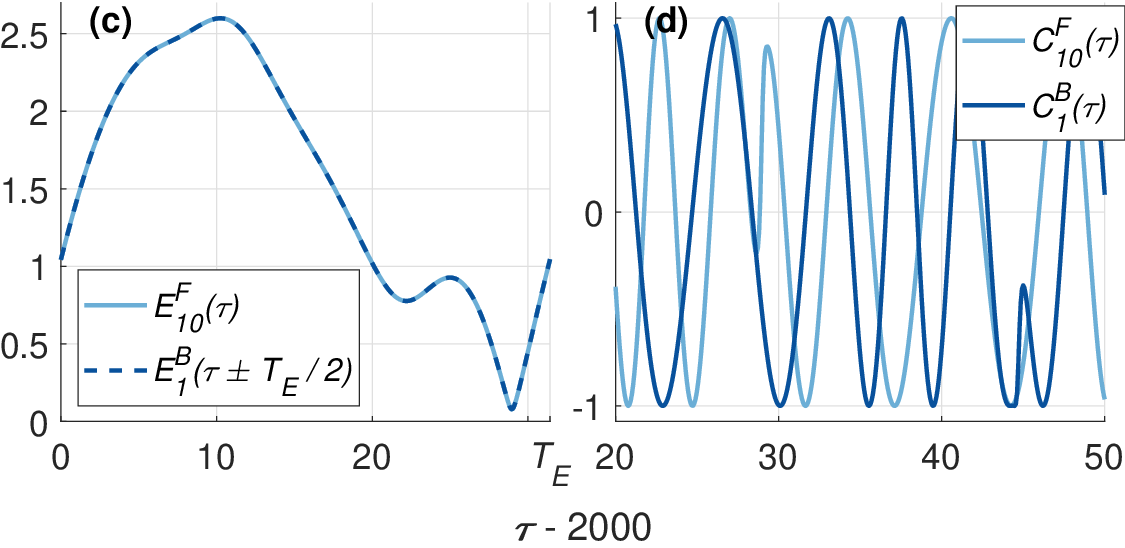}}
\caption{Displacement outputs (a,b), shifted response envelopes (c) and carrier waves (d) calculated with parameters: $n=10$, $\Omega_m=0.2$, $\phi=\pi$, $\Omega_f=0.89$, $K_c=0.7$, $K_m=0.8$, $\zeta=0.02$, $P=1$ and $\mathcal{F}=7$.}
\label{fig_special1}
\end{figure}

A trivial case of nonreciprocity with equal transmitted amplitudes in the opposite directions occurs when $\phi=\pi$ and $n$ is an even number~\cite{paper1}. In this scenario, $|y_{n;q}^F|=|y_{1;q}^B|$ for any integer $q$, regardless of the values of other system parameters. Moreover, $\psi_{n;q}^F = \psi_{1;q}^B$ when $q$ is an even number and $\psi_{n;q}^F = \psi_{1;q}^B \pm \pi$ when $q$ is an odd number. This leads to the following relation between the two response envelopes: $E_n^F(\tau)=E_1^B(\tau \pm T_E/2)$. In other words, the two response envelopes have the same shape with a temporal shift equal to half a period. 
Therefore, we can obtain response that is characterized by nonreciprocal phase shifts with the same envelope shapes without applying the methodology described in Section~\ref{sec:nonrecpPhase}. More importantly, this special case of nonreciprocal phase shifts can be realized at any strength of modulation and for a system of arbitrary length. 

Fig.~\ref{fig_special1} shows the displacement outputs $x(\tau)$, the response envelopes $E(\tau)$ and the carrier waves $C(\tau)$ for a strongly modulated system ($K_m=0.8$) with $\phi=\pi$ and $n=10$; recall Eq.~\eqref{eq_resp2}. We use $\mathcal{F}=7$ in Eq.~\eqref{eq_resp1} to approximate the response accurately. Panels (a,b) show that the anharmonic envelopes of the output displacements are captured well. Panel (c) shows that the plots of $E^F(\tau)$ and $E^B(\tau \pm T_E/2)$ coincide, confirming the half-period phase shift between the two envelopes. As shown in panel (d), the carrier waves are no longer periodic in strongly modulated systems, in contrast to the harmonic carrier waves of weakly modulated systems; {\it cf.} Fig.~\ref{fig_R-example}. Thus, nonreciprocity in strongly modulated systems manifests in both the envelopes and carrier waves. 

\subsection{Near-reciprocal transmission}
\label{sec:special_2}
The constraints in Eq.~(\ref{eq_Eshape}) are applied to the response envelopes and do not directly control the properties of the carrier waves. Interestingly, we have observed that enforcing the constraint for the same envelope shapes can occasionally result in a near-reciprocal response. 

\begin{figure}[b]
\centerline{\includegraphics[scale=0.35]{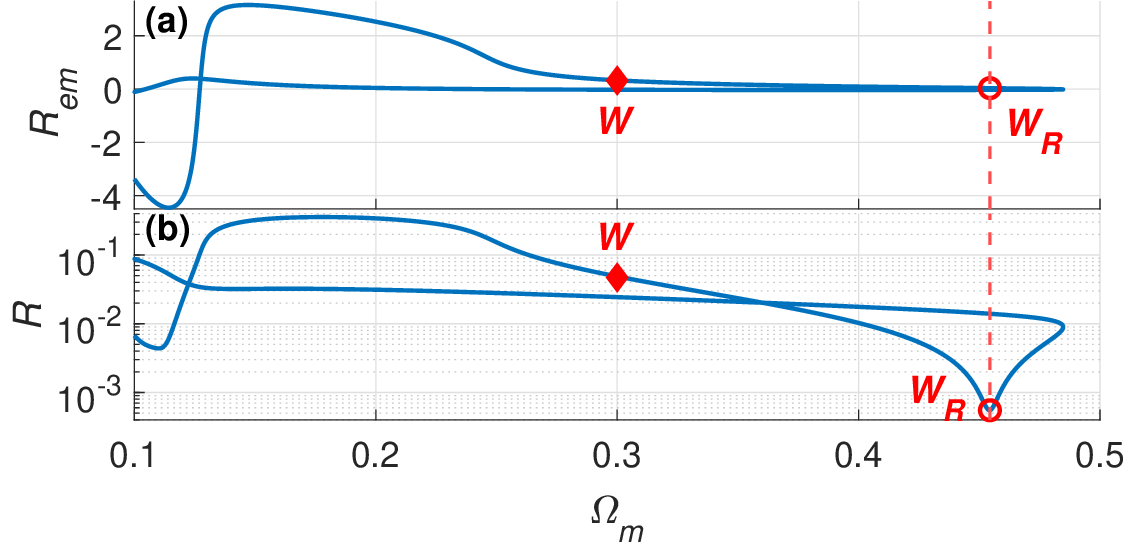}}
\centerline{\includegraphics[scale=0.35]{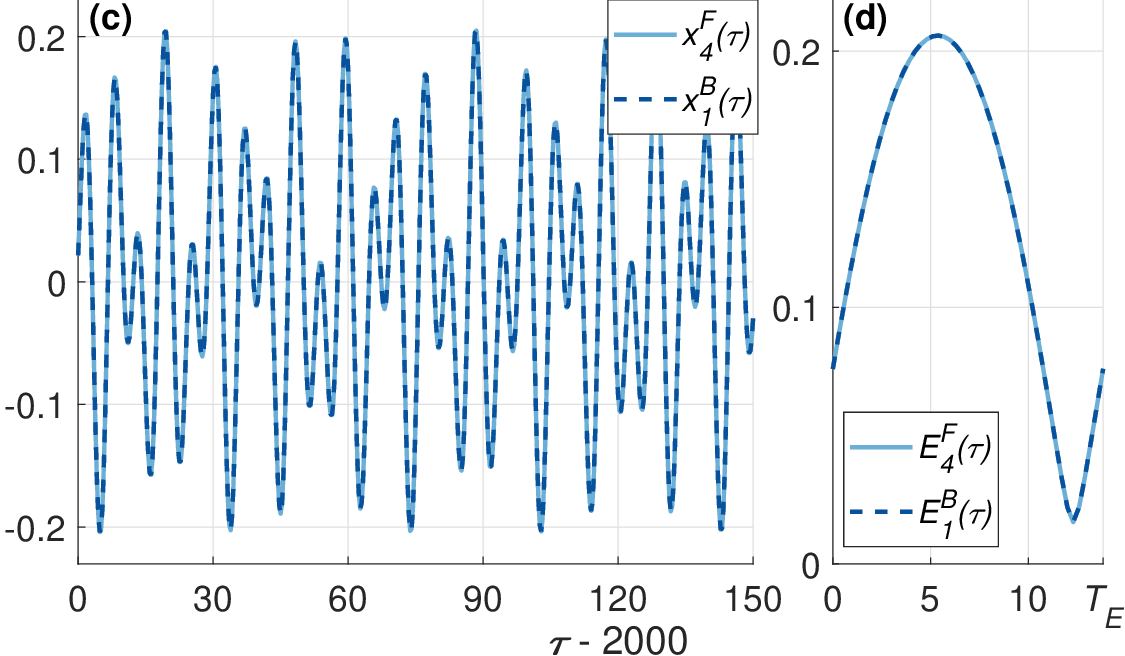}}
\caption{Variation of $\mathcal{R}_{em}$ (a) and $R$ (b) as functions of $\Omega_m$ for point $W$ that exhibits nonreciprocal phase shift. Displacement outputs (c) and response envelopes (d) for point $W_R$.}
\label{fig_special2}
\end{figure}

Fig.~\ref{fig_special2}(a) shows the variation of $\mathcal{R}_{em}$ as a function of $\Omega_m$ for point $W$, representing a set of system parameters that satisfy the constraints in Eqs.~\eqref{eq_PhaseNon} for nonreciprocal phase shifts: $\Omega_m=0.3$, $\Omega_f=0.753$, $\phi=0.703\pi$, $K_c=0.415$, $K_m=0.1$, $n=4$, $\zeta=0.02$ and $P=1$. In Fig.~\ref{fig_special2}(a), $\mathcal{R}_{em} \approx 0$ over a large portion of the curve, especially starting from the portion leading to the turning point near $\Omega_m\approx0.48$ and extending to $\Omega_m\approx0.2$. The {\it forward} and {\it backward} response envelopes have very similar shapes throughout this range of $\Omega_m$. Fig.~\ref{fig_special2}(b) shows the variation of the reciprocity bias, $R$, along the same locus. We have $R<0.04$ throughout the range of $\Omega_m$ where $\mathcal{R}_{em} \simeq 0$, implying that the degree of nonrecpirocity is small. The minimum value of $R$, indicated by point $W_R$, occurs near a zero crossing of $\mathcal{R}_{em}$ for $\Omega_m=0.454$, $\Omega_f=0.638$, $\phi=0.715\pi$ and $K_c=0.713$. We have $R=5.52 \times 10^{-4}$ at this point, 
which means that the two response envelopes almost coincide. Figs.~\ref{fig_special2}(c,d) show the output displacements and response envelopes at point $W_R$. Neither the two output displacements nor the two response envelopes can be visibly distinguished. Vibration transmission is therefore nearly reciprocal for this set of system parameters. 

The near-reciprocal response reported here is reminiscent of the restoring of reciprocity in nonlinear systems with broken mirror symmetry~\cite{giraldo,nonlinear2024}. While restoring reciprocity in nonlinear systems, the reciprocity bias can be made arbitrarily small to ensure reciprocity~\cite{giraldo}. We do not expect the same property to hold in the present work because there is only one symmetry-breaking parameter in our system. Nevertheless, given the small value of $R$ for the example in Fig.~\ref{fig_special2}, it is unlikely that the output displacements could be distinguished in an experimental realization of the system in this case. 

\section{Limitations of the Methodology}
\label{sec:limitations}
\subsection{Systems with $\Omega_f<\Omega_m$}

In analogy to audio communication systems~\cite[Ch. 3]{Commu_systems}, the response envelopes can be viewed as the message signals in the amplitude modulation (AM) technique and the output displacement as the resultant waves after AM. A successful AM process requires that the carrier wave has a much higher frequency than the message signal. Upon receiving the signal, the frequency contents of the signal (peaks at equally distanced frequencies) are used during the demodulation process to retrieve the original message signal. When the frequency of the carrier waves is high enough, the receiver can obtain sufficient peaks within every period of the message signal to reconstruct the original message with high quality. Similarly, if $\Omega_f<\Omega_m$ in our system, there will be less than one intersection point of output displacement and its response envelope within a period $T_E=2\pi/\omega_m$. Thus, the response envelope fails to capture the profile of the output displacement.

In Section~\ref{sec:method_PhaseNon}, we used parameter sets with $0.5\le\Omega_f\le2$ and $\Omega_m=0.2$. The frequency of the response envelopes was therefore lower than the frequencies of the carrier waves, and Eq.~(\ref{eq_envlp}) captured the response envelopes well, as a result; recall Fig.~\ref{fig_output}. In Section~\ref{sec:sameshape}, $\Omega_m$ was allowed to vary as a free parameter to satisfy the additional constraint on envelope shapes, $\mathcal{R}_{em}=0$ in Eq.~\eqref{eq_Eshape}. It is possible that the additional constraint is sometimes satisfied with $\Omega_m>\Omega_f$. Here, we present an example of this scenario that leads to inaccurate prediction of nonreciprocal phase shifts. 

\begin{figure}[htb]
\centerline{\includegraphics[scale=0.35]{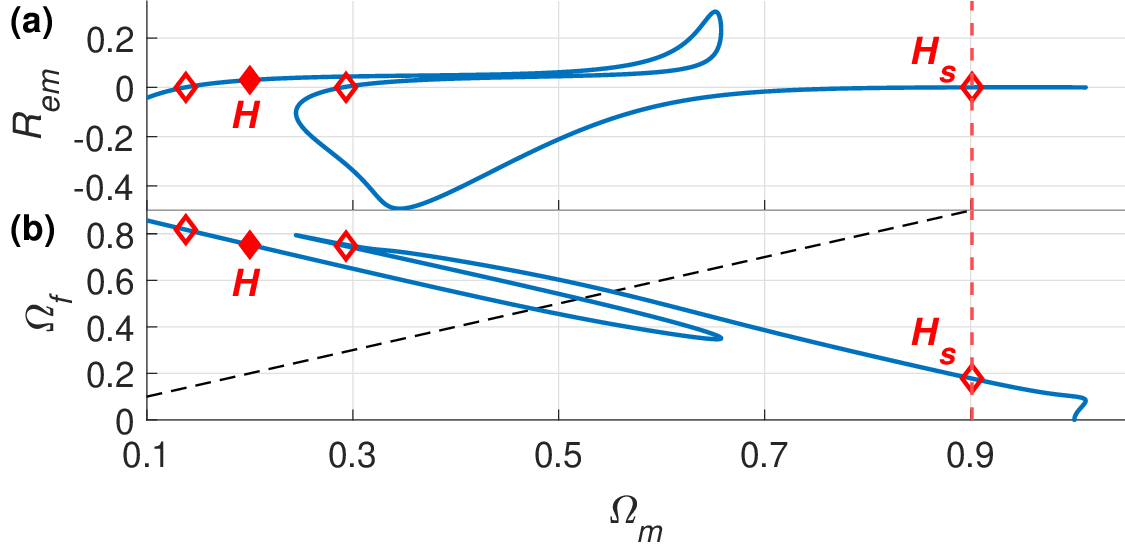}}
\centerline{\includegraphics[scale=0.35]{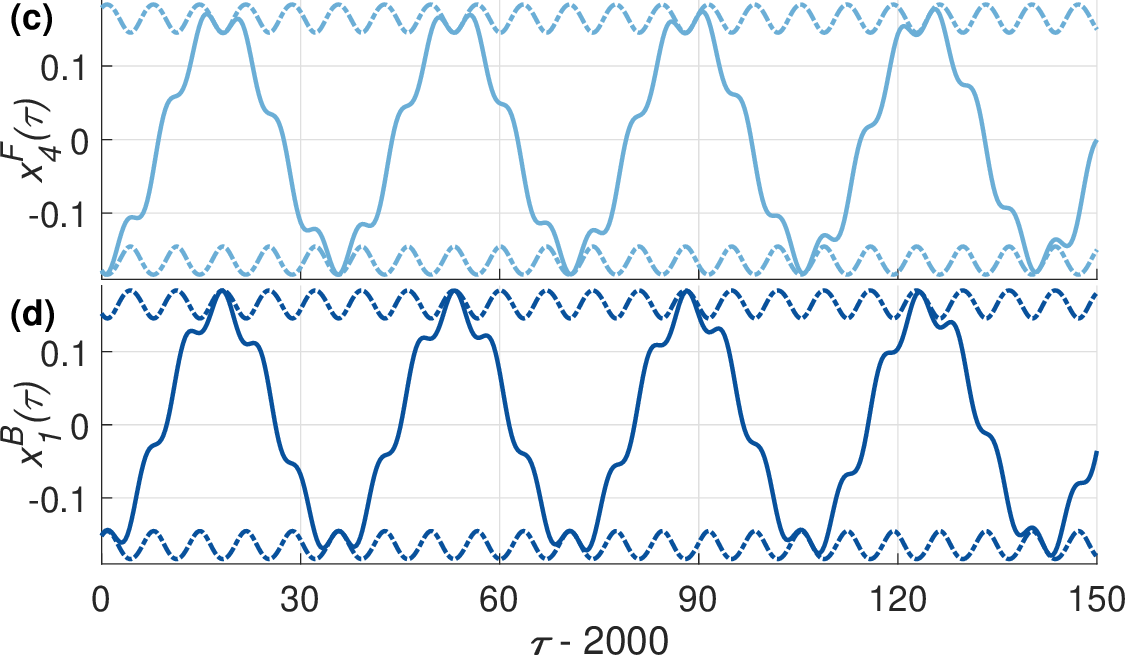}}
\caption{Variation of $\mathcal{R}_{em}$ (a) and $\Omega_f$ (b) as functions of $\Omega_m$ for point $H$ that exhibits nonreciprocal phase shift. The black dashed line in panel (b) represent $\Omega_f=\Omega_m$. (c,d) Displacement outputs and response envelopes for point $H_s$.}
\label{fig_limit}
\end{figure}

Fig.~\ref{fig_limit}(a) shows the variation of $\mathcal{R}_{em}$ as a function of $\Omega_m$ for a set of parameters that satisfies the constraints in Eqs.~(\ref{eq_PhaseNon}) for nonreciprocal phase shifts: $\Omega_m=0.2$, $\Omega_f=0.753$, $\phi=0.458\pi$, $K_c=0.969$, $K_m=0.1$, $n=4$, $\zeta=0.02$ and $P=1$. We refer to this initial point as $H$. As $\Omega_m$ varies, $\mathcal{R}_{em}$ has three zero crossings, indicated by the empty diamond markers. 
Fig.~\ref{fig_limit}(b) shows that $\Omega_m<\Omega_f$ for the first two zero crossings; they fall above the oblique dashed line that indicates $\Omega_m=\Omega_f$. As expected, the response at these two points exhibits nonreciprocal phase shift with the same envelope shape (not shown). However, $\Omega_m>\Omega_f$ at the third zero crossing, indicated by point $H_s$ where $\Omega_m=0.901$, $\Omega_f=0.177$, $\phi=0.473\pi$ and $K_c=4.747$. Figs.~\ref{fig_limit}(c,d) show the output displacements and response envelopes for point $H_s$. The predicted response envelopes have low-amplitude fluctuations of high frequency with large DC shifts ($\mathcal{S}_{dc;4}^F$, $\mathcal{S}_{dc;1}^B$) while the displacements have high-amplitude fluctuations of low frequency. The predicted envelopes do not capture the actual envelope of the response because $\Omega_f<\Omega_m$. 

\subsection{Long systems with strong modulation}
\label{sec:limit2}
Except for the special case presented in Section~\ref{sec:special_1}, all the calculations for nonreciprocal phase shifts in this work are conducted with $\mathcal{F}=1$. We have the same number of free system parameters in this case as the constraint equations, which makes the search for nonreciprocal phase shifts feasible. Nevertheless, the assumption of $\mathcal{F}=1$ restricts the analysis to short, weakly modulated systems~\cite{paper1}. 

A higher value of $\mathcal{F}$ is required to accurately capture the steady-state response envelopes in systems with $K_m>0.1$ or $n>5$. To update the formulation of response envelopes from Section~\ref{sec:envelopes} to a general value of $\mathcal{F}$, we need to use:
\begin{equation}
\label{eq_limF}
\begin{array}{c}
E_{n,1}^{F,B}(\tau) = \sqrt{\mathcal{S}_{dc;n,1}^{F,B}+\mathcal{S}_{ac;n,1}^{F,B}(\tau)},\vspace{2mm} \\
\mathcal{S}_{dc;n,1}^{F,B} = 4 \sum^{\mathcal{F}}_{q=-\mathcal{F}} \big|y_{n,1;q}^{F,B}\big|^2,\vspace{2mm} \\
\mathcal{S}_{ac;n,1}^{F,B}(\tau) = \sum^{2\mathcal{F}}_{j=1} \mathcal{E}_{j;n,1}^{F,B}\cos{(j\Omega_m\tau-\vartheta_{j;n,1}^{F,B})},
\end{array}
\end{equation}
which replaces Eqs.~(\ref{eq_envlp2}-\ref{eq_S_ac}). The following constraints replace Eq.~\eqref{eq_PhaseNon} to enforce a nonreciprocal phase shift between the response envelopes:
\begin{equation}
\label{eq_lim_PhaseNon}
\begin{array}{c}
\mathcal{S}_{dc;n}^F = \mathcal{S}_{dc;1}^B,\vspace{2mm} \\
\mathcal{E}_{1;n}^F = \mathcal{E}_{1;1}^B, \\
\vdots \\
\mathcal{E}_{2\mathcal{F};n}^F = \mathcal{E}_{2\mathcal{F};1}^B.
\end{array}
\end{equation}
There are $2\mathcal{F}+1$ constraints in Eq.~\eqref{eq_lim_PhaseNon}. A larger value of $\mathcal{F}$ increases the number of constraints, but it does not change the number of available system parameters that can be used to satisfy the additional constraints. As discussed in Section~\ref{sec:nonrecpPhase}, only 4 system parameters can be changed as free (control) parameters: $\Omega_f$, $\phi$, $K_c$ and $\Omega_m$. If $\mathcal{F} >1$, the number of equations exceed the number of unknowns (overdetermined system) and it becomes no longer possible to satisfy all the constraints in Eq.~\eqref{eq_lim_PhaseNon}. Thus, the methodology introduced in Section~\ref{sec:nonrecpPhase}, in its current form, is limited to $\mathcal{F}=1$, which corresponds to short systems ($n<5$) subject to weak modulations ($K_m \leq 0.1$). 

This is not an insurmountable limitation, however. Within the framework of the present work, one potential way to overcome this limitation is to allow more system parameters to vary (increasing the number of free parameters). Currently, parameter $\phi$ makes the only difference between the units. It is possible to allow for variation in the system parameters across units (mass, stiffness, etc.) to make up for the required number of control parameters. 
Alternatively, it may be possible to develop an alternative formulation to find nonreciprocal phase shifts in strongly modulated systems with more number of units. 


\section{Conclusions}
\label{sec:conclusion}
We reported on the existence of response regimes in spatiotemporally modulated systems that are characterized by nonreciprocal phase shifts. This is a special case of nonreciprocal dynamics in which the transmitted vibrations have the same amplitude and the only contributor to nonreciprocity is the difference between the transmitted phases in opposite directions. This attribute of nonreciprocity is rarely discussed in the context of spatiotemporally modulated systems. 

We presented a methodology for obtaining nonreciprocal phase shifts that takes advantage of the time-periodic nature of the envelopes of the response in the steady state. This circumvents the complexities of the non-periodic nature of the response caused by the presence of two incommensurate frequencies (modulation and external drive). While we primarily focused on weakly modulated systems with a small number of units, we also presented a special case of nonreciprocal phase shifts in a system of arbitrary length and strength of modulation. In addition, we provided a formulation that ensures the same transmitted waveforms in opposite directions, which also helped us obtain a special case of near-reciprocal transmission of vibrations. 

In summary, we extended the phenomenon of phase nonreciprocity from nonlinear systems (passive) to spatiotemporally modulated systems (active). We discussed the main limitations of our methodology in its current form: systems with modulation that is faster than the external drive, and long systems subject to strong modulation. We point out a potential way to overcome the second limitation to motivate further research on the topic. We hope that our findings can enable new developments in wave processing techniques such as phase shift keying and communication devices. 

\begin{acknowledgments}
We are thankful to H.M. Osinga, University of Auckland, for a very helpful discussion on systems of algebraic equations. We acknowledge financial support from the Natural Sciences and Engineering Research Council of Canada through the Discovery Grant program. J.W. acknowledges additional support from Concordia University and from Centre de Recherches Mathématique, Quebec.
\end{acknowledgments}

\appendix

\section{Non-dimensionalization} 
\label{appendix:nondimensionalization}

The equations of motion which govern the $n$-DoF modulated system in Fig.~\ref{fig_nDoF} are:
\begin{equation}
\label{eqA1}
\begin{array}{c}
\displaystyle m \dv[2]{u_1}{t} + c \dv{u_1}{t} + k_1 u_1 
+ k_c \delta^2_1 = F_1 \cos{(\omega_f t)},\\
\vdots \\
\displaystyle m \dv[2]{u_p}{t} + c \dv{u_p}{t} + k_p u_p 
+ k_c \delta^2_p = 0,\\
\vdots \\
\displaystyle m \dv[2]{u_n}{t} + c \dv{u_n}{t} + k_n u_n 
+ k_c \delta^2_n = F_n \cos{(\omega_f t)},
\end{array}
\end{equation}
where $k_p=k_{g,DC}+k_{g,AC}\cos(\omega_m t - \phi_p)$ and $\phi_p=(p-1)\phi$ ($p=1,2,\cdots,n$). The difference term $\delta^2_p=2u_p-u_{p-1}-u_{p+1}$ everywhere, except at the two ends where $\delta^2_1=u_1-u_2$ and $\delta^2_n=u_n-u_{n-1}$.
We use $\tau=\omega_0t$ as the nondimensional time with $\omega_0=\sqrt{k_{g,DC}/m}$.
We define $\zeta=c/(2m\omega_0)$, $\Omega_m=\omega_m/\omega_0$, $\Omega_f=\omega_f/\omega_0$,  $K_c=k_c/k_{g,DC}$, $K_m=k_{g,AC}/k_{g,DC}$, $P_1=F_1/(ak_{g,DC})$, $P_n=F_n/(ak_{g,DC})$ and $x_p=u_p/a$, where $a$ is a representative length.
After substituting these parameters into \eqN{eqA1}, we obtain:
\begin{widetext}
\begin{equation}
\label{eqA2}
\begin{array}{c}
\displaystyle m a \omega_0^2 \ddot{x}_1 +2\zeta m a \omega_0^2 \dot{x}_1 + k_{g,DC} a x_1 \left[1 + K_m \cos{\left( \Omega_m \tau \right)} \right] 
+ K_c k_{g,DC} a \left( x_1 - x_2 \right) = P_1 a k_{g,DC} \cos{\left( \Omega_f \tau \right)},\\
\vdots \\
\displaystyle m a \omega_0^2 \ddot{x}_p +2\zeta m a \omega_0^2 \dot{x}_p + k_{g,DC} a x_p \left[1 + K_m \cos{\left( \Omega_m \tau - \phi_p \right)} \right] 
+ K_c k_{g,DC} a \left( 2 x_p - x_{p+1} - x_{p-1} \right) = 0,\\
\vdots \\
\displaystyle m a \omega_0^2 \ddot{x}_n \!+\! 2\zeta m a \omega_0^2 \dot{x}_n \!+\! k_{g,DC} a x_n \left[1 \!+\! K_m \cos{\left( \Omega_m \tau \!-\! \phi_n \right)} \right] 
\!+\! K_c k_{g,DC} a \left( x_n \!-\! x_{n-1} \right) \!=\! P_n a k_{g,DC} \cos{\left( \Omega_f \tau \right)},
\end{array}
\end{equation}
\end{widetext}
where $\ddot{x}_p$ and $\dot{x}_p$ represent $\dd[2]{x_p} / \dd[2]{\tau}$ and $\dd{x_p} / \dd{\tau}$ respectively.
\eqN{eqA2} can be further simplified as:
\begin{equation}
\label{eqA3}
\begin{array}{c}
\displaystyle \ddot{x}_1 +2\zeta \dot{x}_1 + x_1 \left[1 + K_m \cos{\left( \Omega_m \tau \right)} \right] \\ 
\displaystyle + K_c \left( x_1 - x_2 \right) = P_1 \cos{\left( \Omega_f \tau \right)},\\
\vdots \\
\displaystyle \ddot{x}_p +2\zeta \dot{x}_p + x_p \left[1 + K_m \cos{\left( \Omega_m \tau - \phi_p \right)} \right] \\ 
\displaystyle + K_c \left( 2 x_p - x_{p+1} - x_{p-1} \right) = 0,\\
\vdots \\
\displaystyle \ddot{x}_n +2\zeta \dot{x}_n + x_n \left[1 + K_m \cos{\left( \Omega_m \tau - \phi_n \right)} \right] \\ 
\displaystyle + K_c \left( x_n - x_{n-1} \right) = P_n \cos{\left( \Omega_f \tau \right)},
\end{array}
\end{equation}

In this paper, calculations and analysis of the response of the $n$-DoF modulated system are all based on \eqN{eqA3}, which is the same as \eqN{eq_EoM_p}.

\section{Response envelope: Amplitude terms}
\label{appendix:envelopeAB}

The envelope amplitudes ($\mathcal{A}_p$ and $\mathcal{B}_p$) and envelope phases ($\theta_{a;p}$ and $\theta_{b;p}$) in Eq.~(\ref{eq_S_ac}) can be calculated from:
\begin{align*}
\mathcal{A}_p =& 8\sqrt{\left(Q_{c;a;p}\right)^2+\left(Q_{s;a;p}\right)^2},\nonumber \\
\theta_{a;p} =& \atan\!2\left(Q_{s;a;p},Q_{c;a;p}\right), \nonumber \\
\mathcal{B}_p=& 8\sqrt{\left(Q_{c;b;p}\right)^2+\left(Q_{s;b;p}\right)^2},\nonumber \\
\theta_{b;p}=& \atan\!2\left(Q_{s;b;p},Q_{c;b;p}\right),
\end{align*}
where $Q_{c,s;a,b;p}$ are defined as:
\begin{align*}
Q_{c;a;p} =& \Re(y_{p,0})\left(\Re(y_{p,-1})+\Re(y_{p,1})\right) \nonumber \\
&+\Im(y_{p,0})\left(\Im(y_{p,-1})+\Im(y_{p,1})\right), \nonumber \\
Q_{s;a;p} =& \Re(y_{p,0})\left(\Im(y_{p,-1})-\Im(y_{p,1})\right) \nonumber \\
&-\Im(y_{p,0})\left(\Re(y_{p,-1})-\Re(y_{p,1})\right), \nonumber \\
Q_{c;b;p} =& \Re(y_{p,-1})\Re(y_{p,1})+\Im(y_{p,-1})\Im(y_{p,1}), \nonumber \\
Q_{s;b;p} =& \Re(y_{p,1})\Im(y_{p,-1})-\Re(y_{p,-1})\Im(y_{p,1}). \nonumber
\end{align*}

\bibliography{apssamp}

\end{document}